\documentclass[twocolumn,amsmath,amssymb,longbibliography,noeprint]{revtex4-1}

\usepackage{bbm}
\usepackage[caption=false]{subfig}
\usepackage{overpic}
\usepackage{amsmath}
\usepackage{epsfig,psfrag}
\usepackage{pstricks}
\usepackage{dcolumn}
\usepackage{bm}
\usepackage{graphicx}
\usepackage{color}
\usepackage{version}
\usepackage{hyperref}
\definecolor{darkblue}{RGB}{46,48,146}
\hypersetup{
   colorlinks=true,
   citecolor={darkblue},
   linkcolor={darkblue},
   urlcolor={darkblue},
   pdfhighlight=/I
}

\usepackage{times}
\usepackage{xcolor}
\input{Qcircuit}

\begin{document}
\title{Combining Topological Hardware and Topological Software: \\Color Code Quantum Computing with Topological Superconductor Networks}
\author{Daniel Litinski, Markus S.\ Kesselring, Jens Eisert, and Felix von Oppen}
\affiliation{Dahlem Center for Complex Quantum Systems and Fachbereich Physik, Freie Universit\"at Berlin, Arnimallee 14, 14195 Berlin, Germany}

\begin{abstract}
We present a scalable architecture for fault-tolerant topological quantum computation using networks of voltage-controlled Majorana Cooper pair boxes, and topological color codes for error correction. Color codes have a set of transversal gates which coincides with the set of topologically protected gates in Majorana-based systems, namely the Clifford gates. In this way, we establish color codes as providing a natural setting in which
 advantages offered by topological hardware can be combined with those arising from  topological error-correcting software
 for full-fledged fault-tolerant quantum computing. We 
 provide a complete description of our architecture including 
the underlying physical ingredients. We start by showing that in topological superconductor networks, hexagonal cells can be employed to serve as physical qubits for universal quantum computation, and present protocols for realizing
topologically protected Clifford gates. These hexagonal cell qubits allow for a direct implementation of open-boundary color codes with ancilla-free syndrome readout and logical $T$ gates via magic state distillation. For concreteness, we describe how the necessary operations can be implemented using networks of Majorana Cooper pair boxes, and give a feasibility estimate for error correction in this architecture. Our approach is motivated by nanowire-based networks of topological superconductors, but could also be realized in alternative settings such as quantum Hall-superconductor hybrids.
\end{abstract}

\maketitle

\section{Introduction}\label{sec:intro}

\begin{figure*}[t]
\centering
\def\svgwidth{\linewidth}
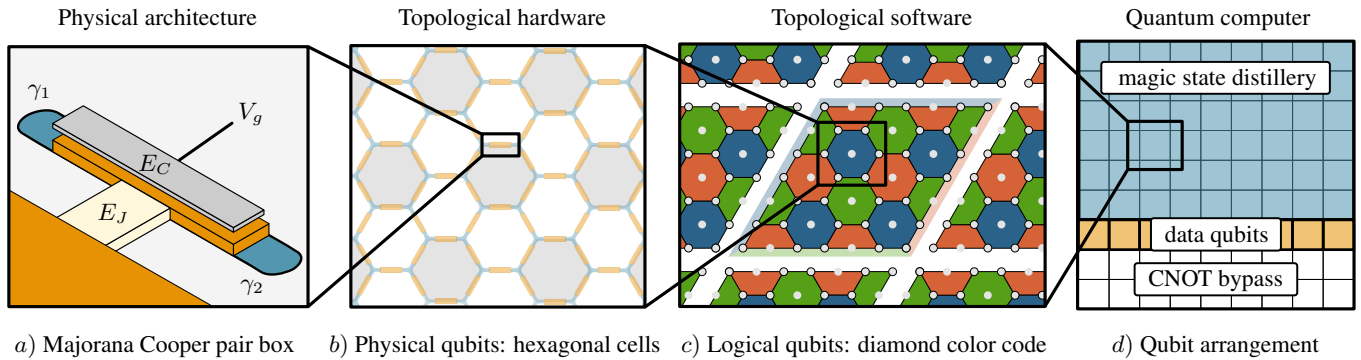
\caption{Overview of the design for a scalable fault-tolerant topological quantum computer presented in this article. The basic building block is the Majorana Cooper pair box $(a)$ consisting of a topological superconducting island with charging energy $E_C$ and Josephson energy $E_J$ hosting a pair of Majoranas $\gamma_1$ and $\gamma_2$. Parity measurements of the island are controlled by a gate voltage $V_g$. Multiple connected Majorana Cooper pair boxes form a topological superconductor network through which Majoranas can be moved, and which allows for the measurement of $2n$-Majorana parity operators. A triangular lattice of hexagonal cell qubits $(b)$ allows for universal quantum computation with topologically protected Clifford gates. Fault-tolerance is added by encoding hexagonal cell qubits in diamond color codes $(c)$ with transversal Clifford gates. These form a square lattice of logical qubits. Arranging qubits on a line $(d)$ with a magic state distillery and a CNOT bypass completes the universal gate set with a logical $T$ gate, and allows for CNOT gates between any pair of data qubits with constant time overhead.}
\label{fig:intro}
\end{figure*}

Physical realizations of large-scale quantum computers remain a paramount
experimental challenge due to the unavoidable presence of environmental decoherence. Topological quantum computing 
is generally seen as paving the way towards a solution of this problem~\cite{Kitaev2003,Nayak2008,TerhalRMP}. In fact, this is expected to be 
true in more than one sense: In a mindset of 
condensed matter physics, excitations of topological phases of matter have been identified as 
candidates for physical qubits that are robust to local perturbations, and on which a certain set of quantum gate operations 
can be performed largely noise-free. In the context of 
quantum information theory, topological quantum error-correcting codes have been devised as codes
featuring high error tolerance which only require the measurement of local stabilizer operators. While clearly
related, these predominantly  hardware-based and software-based approaches constitute two distinctly different readings
of topological quantum computing.

On the hardware side, the interplay of superconductivity, spin-orbit coupling and single spin-polarized conducting channels has inspired various proposals for experimental realizations of \textit{Majorana zero modes}~\cite{Fu2008,Oreg2010,Lutchyn2010,Alicea2012,Leijnse2012,Beenakker2013}, subsequently simply referred to as \textit{Majoranas}. 
Quantum information can be encoded using spatially separated pairs of Majoranas~\cite{Kitaev2001} whose parity state is unaffected by local perturbations. We refer to qubits encoded using this parity state as \textit{physical qubits} arising from \textit{topological hardware}. Furthermore, the exchange of pairs of Majoranas constitutes a nontrivial braiding operation which can be used for the implementation of robust quantum gates. Recent experiments have provided increasing evidence for the emergence of Majorana zero modes in semiconducting nanowires with mesoscopic superconducting islands~\cite{Mourik2012,Das2012,Albrecht2016a,Deng2016,Suominen2017}.
In such setups, the state of the Majorana pair depends on the fermion parity of the mesoscopic island. Therefore, electrons tunneling onto the island can change the parity state and thus spoil any quantum information encoded by the Majoranas. This process is called quasiparticle poisoning. Among other error sources, its rate defines 
a finite lifetime for Majorana-based qubits.

If one aims at storing and manipulating quantum information beyond the quasiparticle poisoning time~--~in fact, in principle, for
arbitrary times~--~errors need to be actively corrected. This can be achieved by making use of \textit{topological error-correcting codes}. The basic principle of such codes is to fight local errors with entanglement, so that local noise cannot affect the logical information~\cite{Preskill1998}. This is done by using multiple physical qubits to encode a single \textit{logical qubit}, which we refer to as \textit{topological software}.  

Recent proposals~\cite{TerhalMajorana,Vijay2015,Vijay2016,Landau2016,Plugge2016} 
have taken key steps in the direction of combining topological hardware with software. Importantly, 
the combination of Majorana-based qubits with topological surface codes has been studied. Yet, while the replacement of physical qubits with logical qubits enhances resilience against noise, one must be aware that one also substitutes physical gates with logical gates. Topological protection of gates on the physical level does not necessarily translate to logical gates, since in general, physical and logical gates are unrelated. Any error-correcting code is in principle allowed to have a (non-universal~\cite{Eastin2008}) set of transversal gates~--~i.e., \setcitestyle{super} logical gates that correspond to the simultaneous application of the same physical gate to all physical qubits~\footnote{While this is the notion of transversal gates that we use in this paper, the general definition of transversal gates includes all operations that are tensor products of operations on local subsystems.}~--~but the transversal gates of surface codes are limited to the CNOT and Pauli gates.

\setcitestyle{numbers}

In this work, we go a significant step further and establish two-dimensional
topological color codes as a natural fit to enhance the fault tolerance of Majorana-based quantum computers. They seamlessly combine the fermion parity-protected topological order of topological superconductors~\cite{Bonderson2013} with the long-range topological order of the toric code~\cite{Bombin2012} in a way that allows one to exploit the topological protection of both. Compared to surface codes, color codes not only have a richer set of transversal gates, but this set also coincides with the  gates that are accessible by braiding of Majoranas, namely the Clifford gates~\cite{Bombin2006,Landahl2011}. We hence
further contribute to identifying the precise advantages offered by topological protection both as far as the underlying condensed
matter physics is concerned, as well as on the level of logical encoding.

In the following sections, we describe our design for a scalable fault-tolerant topological quantum computer from the ground up, discussing the microscopic details of the Majorana-based physical qubits, their encoding in topological superconductor networks, and the arrangement and manipulation of logical qubits for quantum computing (see Fig.~\ref{fig:intro}). We begin in Sec.~\ref{sec:physqubits} by describing how networks of topological superconducting islands can be used for universal quantum computation with topologically protected Clifford gates. We require topological superconductor networks to be capable of three operations: Moving Majoranas through the network by coupling neighboring islands, measuring $2n$-Majorana parity operators on connected islands, and lifting the degeneracy of the parity states on an island. 
We show that in such networks, physical qubits can be arranged in hexagonal cells with six nearest neighbors, such that the qubits form a triangular lattice (gray hexagons in Fig.~\ref{fig:intro}b). Here, each hexagonal cell is associated with four Majoranas that are used for quantum computation. Universal quantum computation requires the implementation of a universal set of quantum gates. One such set consists of the Clifford gates (Hadamard, $\pi/4$ and CNOT gate) and the $T$ gate (or $\pi/8$ gate). We present protocols for single-qubit Clifford gates via braiding inside a hexagonal cell, and CNOT gates between any pair of cells via braiding and parity measurements. The addition of an unprotected $T$ gate~--~which is not accessible via braiding of Majoranas~--~by controlled splitting of the degeneracy completes the universal gate set.

While the Clifford gates of these Majorana-based qubits are topologically protected, the $T$ gate requires fine-tuning of the device control parameters, which can easily lead to errors in the $T$ gate. Instead of attempting to implement a robust $T$ gate on the level of physical qubits~\cite{Karzig2015,Barkeshli2015}, we address this problem using magic state distillation. This is a common proposal for a fault-tolerant implementation of the $T$ gate on the level of logical qubits, the precision of which scales with the protocol length~\cite{Bravyi2005}. These protocols typically include many multi-target CNOT gates (i.e., multiple CNOT gates with the same control but different target qubits). We show how parity measurements in topological superconductor networks can be used for fast multi-target CNOT gates, replacing multiple CNOT gates by a protocol that is as fast as a single CNOT.

Another advantage of Majorana-based qubits is ancilla-free syndrome readout. Quantum error-correcting codes typically require the measurement of stabilizer operators of the form $\sigma_z^{\otimes n}$, where $\sigma_z$ is a Pauli matrix and $n$ is the number of qubits involved in the measurement. In conventional setups for quantum computing, such $n$-qubit parity operators are typically not directly measurable, but require a lengthy protocol involving an ancilla qubit and $n$ CNOT gates. Since in topological superconductor networks, the parity of $2n$ Majoranas can be measured directly if the Majoranas are moved onto a single connected superconducting island, $n$-qubit parity operators can be measured without the use of ancilla qubits. In preparation for the color code, we demonstrate how hexagonal cell qubits can be used to measure the required 6-qubit parity operators.

The triangular lattice of physical qubits allows for a direct implementation of triangular color codes with transversal Clifford gates. In contrast to fermionic codes~\cite{Bravyi2010,Vijay2015,Vijay2017,Hastings2017} where each lattice site corresponds to a Majorana fermion, we use a bosonic code where each lattice site is a bosonic degree of freedom, since our physical qubits are comprised of \textit{four} Majorana fermions each, and are therefore bosonic qubits.
In our encoding scheme, the logical qubits are arranged on a square lattice, where each logical qubit has four nearest neighbors. As this leaves some unused hexagonal cells, we extend the triangular color codes to diamond-shaped color codes (see Fig.~\ref{fig:intro}c), which have the same code distance as their triangular counterparts, but a lower logical error rate. In Sec.~\ref{sec:logqubits}, we show that a square arrangement of diamond color code qubits (see Fig.~\ref{fig:intro}d) can be used for universal fault-tolerant quantum computing with topologically protected Clifford gates, constant-time CNOT gates between any pair of logical qubits, and logical $T$ gates with arbitrary precision. We discuss various protocols for logical CNOT and multi-target CNOT gates, based on transversal gates and lattice surgery~\cite{Landahl2014}. 

In order to show a possible scalable realization of topological superconductor networks, we review Majorana Cooper pair boxes~\cite{Vijay2015,Aasen2016,Bouchiat1998,Koch2007,Hassler2011,VanHeck2012,Hyart2013} in Sec.~\ref{sec:hardware} as basic building blocks of the physical architecture. In our description of Majorana Cooper pair boxes (see Fig.~\ref{fig:intro}a), we revisit how topological superconducting islands combined with capacitive coupling via a top gate and Josephson coupling to a bulk superconductor can be used for parity-to-charge conversion~\cite{Aasen2016}. We demonstrate that networks of Majorana Cooper pair boxes are capable of performing the aforementioned required operations. These can be implemented using proximitized semiconductor nanowires, on which recent experiments have focused, but possibly also in other platforms such as hybrid structures based on quantum Hall, quantum spin Hall or quantum anomalous Hall edge states.

Finally, in Sec.~\ref{sec:feasibility}, we consider the main error sources in our physical architecture and give a feasibility estimate. There are three time-scales that characterize networks of Majorana Cooper pair boxes: the time required to move Majoranas, the duration of parity measurements, and the quasiparticle poisoning time. We identify constraints that physical setups need to satisfy in order to operate below the error threshold of color codes. Using a Monte Carlo simulation, we study the improved performance of diamond color codes over triangular color codes, and give an estimate of the space overhead~--~i.e., the number of physical qubits per logical qubit~--~required for the logical qubits to reach sufficiently long survival times for quantum computation on the basis of experimental measurements of quasiparticle poisoning times~\cite{VanWoerkom2015,Higginbotham2015,Albrecht2016}.

It should be clear that this article is aimed at both the condensed matter and quantum information communities. 
Therefore, we have made an effort to include basic introductions to the relevant concepts. Still, this article is by no means a review, but meant to lay the groundwork for color code quantum computing with Majoranas in order to fully exploit the topological protection of Majorana-based qubits.

\section{Topological hardware: \\ hexagonal cell qubits}
\label{sec:physqubits}

\begin{figure}
\centering
\def\svgwidth{\linewidth}
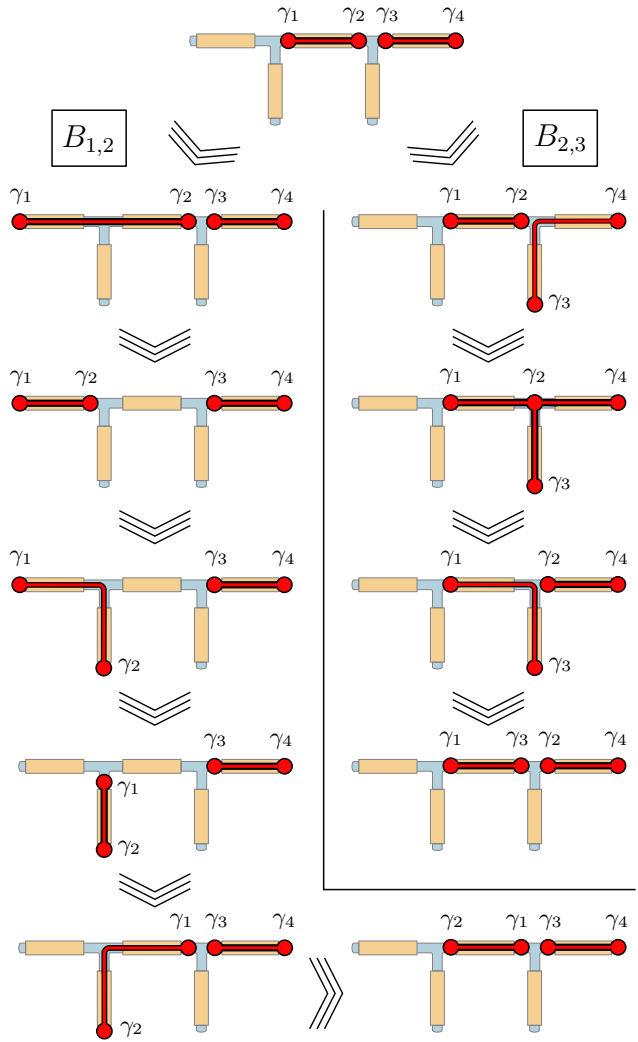
\caption{Protocols for braiding operations in a double T-junction, where red dots denote Majoranas and red lines connect the coupled superconducting islands (orange). Left: Braiding of $\gamma_1$ and $\gamma_2$ is achieved via a three-point turn in the left T-junction. Right: To braid $\gamma_2$ and $\gamma_3$, first $\gamma_3$ is moved from the right island to the bottom right island. Then, $\gamma_2$ is moved to the right island by first connecting all three islands in the right T-junction and then disconnecting the right island. Finally, $\gamma_3$ is moved to the center island.}
\label{fig:doubletjunction}
\end{figure}

\begin{figure}[t]
\centering
\def\svgwidth{\linewidth}
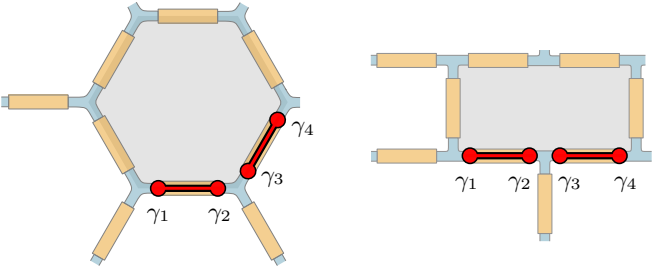
\caption{Left: Hexagonal cell hosting four Majoranas encoding one qubit. The single-qubit Clifford gates can be performed by braiding in the double T-junction in the lower part of the cell. In a network of such cells, each cell has up to six neighbors. Right: Such a hexagonal lattice can also be realized with only two different wire orientations using a brick wall geometry of superconducting islands.}
\label{fig:hexcell}
\end{figure}

In a topological superconductor network, each superconducting island can host a pair of Majoranas $\gamma_1$ and $\gamma_2$ with degenerate even $\ket{e}$ and odd $\ket{o}$ eigenstates of the fermion parity operator $i\gamma_1\gamma_2$. We require the network to be capable of three basic operations:

\begin{enumerate}
	\item Majoranas can be moved from island to island by connecting neighboring superconducting islands (see Fig.~\ref{fig:doubletjunction}). 
	\item For $2n$ Majoranas on a single connected island, the total parity operator $i^n\prod_{j=1}^{2n} \gamma_j$ of $2n$ Majoranas can be measured projectively.
	\item The degeneracy between $\ket{e}$ and $\ket{o}$ can be split temporarily and restored again.
\end{enumerate}

We now show that such networks can be used to realize a universal quantum computer. Even though a pair of Majoranas is a two-level system, no superposition of $\ket{e}$ and $\ket{o}$ can exist due to fermion-parity superselection, and therefore a pair of Majoranas cannot be used as a qubit. Instead, qubits are encoded using two islands hosting four Majoranas with fixed \textit{total} fermion parity (see Fig.\ \ref{fig:doubletjunction}), either in the even parity sector
\begin{equation}
	\ket{0} = \ket{e,e}, \quad \ket{1} = \ket{o,o} \, ,
\label{eqn:qubits1}
\end{equation}
or in the odd parity sector
\begin{equation}
	\ket{0} = \ket{e,o}, \quad \ket{1} = \ket{o,e} \, .
\label{eqn:qubits2}
\end{equation}
To initialize a qubit in one of these states, the two-Majorana fermion parity of both islands is measured.
Both encodings can be used interchangeably, as in both cases the qubit is measured in the computational basis by measuring the parity on the first island.

Furthermore, in both encodings, the exchange of $\gamma_1$ and $\gamma_2$, and of $\gamma_2$ and $\gamma_3$ performs the same braiding operations $B_{1,2}$ and $B_{2,3}$ respectively. Since the braiding operator~\cite{Leijnse2012}
\begin{equation}
	B_{i,j} = \frac{1 + \gamma_i \gamma_j}{\sqrt{2}}
\end{equation}
describes the clockwise exchange of Majoranas $\gamma_i$ and $\gamma_j$, the braiding operators describe the qubit operations
\begin{equation}
	B_{1,2} = e^{-i\frac{\pi}{4}\sigma_z}, \quad B_{2,3} = e^{-i\frac{\pi}{4}\sigma_x} \, .
	\label{eqn:braid}
\end{equation}
Here, $\sigma_z$ and $\sigma_x$ are Pauli operators in the computational basis $\{ \ket{0}, \ket{1} \}$. In terms of Majorana operators, $\sigma_z = i\gamma_1\gamma_2$ and $\sigma_x = i\gamma_2 \gamma_3$.

Universal quantum computation requires a universal set of quantum gates, i.e., a set of unitary operations on the qubits, such that any n-qubit unitary operation can be constructed as a product of unitaries from the universal set. One such universal gate set is the standard set $\{H,T,S,{\rm CNOT} \}$~\cite{Boykin2000}, in which \linebreak $S=\exp(-i\pi \sigma_z / 4)$ and $T=\exp(-i\pi \sigma_z / 8)$ are the $S$ and $T$ gate (equivalently $\pi/4$ and $\pi/8$ gate), and $H$ and CNOT are the Hadamard and controlled-not gate
\begin{equation}
	H = \frac{1}{\sqrt{2}}\begin{pmatrix}
	1 & 1 \\
	1 & -1 
	\end{pmatrix}, \quad
	{\rm CNOT} = \begin{pmatrix}
		1 & 0 & 0 & 0 \\
		0 & 1 & 0 & 0 \\
		0 & 0 & 0 & 1 \\
		0 & 0 & 1 & 0
	\end{pmatrix} \, .
\end{equation}
The gates generated by the non-universal set $\{H, S, {\rm CNOT} \}$ form the set of so-called Clifford gates, which are those gates that map multi-qubit 
Pauli operators to Pauli operators under conjugation.

\subsection{Clifford gates in hexagonal cell qubits}

From Eq.~\eqref{eqn:braid} it is evident that the single-qubit Clifford gates can be implemented by braiding, since $S = B_{1,2}$ and \linebreak $H = i B_{1,2}B_{2,3}B_{1,2} = iB_{2,3}B_{1,2}B_{2,3} $. A topological superconductor network that allows for both exchanges is the double T-junction~\cite{Alicea2011,Aasen2016}. In this five-island geometry, the upper and right superconducting islands host four Majoranas encoding a qubit. Figure~\ref{fig:doubletjunction} shows protocols for the braiding operations $B_{1,2}$ via a three-point turn in the left T-junction, and $B_{2,3}$ using the right T-junction.

In the remainder of this section, we will show that arrays of hexagonal cell qubits depicted in Fig.~\ref{fig:hexcell} can be used for universal quantum computation, where qubits are arranged on a triangular lattice with up to six nearest neighbors for each qubit. Since the lower part of the hexagonal cell is a double T-junction, it can be used for single-qubit Clifford gates by braiding. Note that if the physical implementation only allows for two orientations of the wires, as opposed to three, such hexagonal cells can also be embedded into a lattice with a brick wall geometry, where hexagonal cells are equivalent to nine-island rectangular cells.

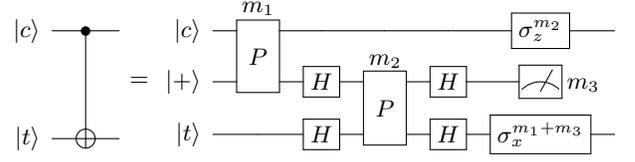
\begin{figure}
\begin{tabular}{ccc}
\scalebox{1}{
\Qcircuit @C=1em @R=4em {
    \lstick{\ket{c}} & \ctrl{1} & \qw \\
    \lstick{\ket{t}} & \targ & \qw 
    }}
&
\raisebox{-2.3em}{$= \quad \quad$}
&
\scalebox{1}{
 \Qcircuit @C=1em @R=.7em {
    \lstick{\ket{c}} & \multigate{1}{P} & \qw & \qw & \qw & \gate{\sigma_z^{m_2}} & \qw \\
    \lstick{\ket{+}} & \ghost{P} &\gate{H} & \multigate{1}{P} & \gate{H} & \meter & \\
    \lstick{\ket{t}} &  \qw & \gate{H} & \ghost{P} & \gate{H} & \gate{\sigma_x^{m_1+m_3}} & \qw
    }}
    \put (-141,7) {$m_1$}
    \put (-93,-13) {$m_2$}
    \put (-18,-22) {$m_3$}
\end{tabular}
\caption{Quantum circuit for a CNOT gate using parity measurements and an ancilla qubit initialized in the $\ket{+}$ state. First, the parity operator $\sigma_z \otimes \sigma_z$ of control and ancilla is measured, with outcome $m_1$. Next, the parity operator $\sigma_x \otimes \sigma_x$ of ancilla and target is measured, with outcome $m_2$. Finally, the ancilla is measured in the computational basis $\sigma_z$ with outcome $m_3$. The three outcomes determine the final correctional operation $\sigma_z^{m_2} \otimes \sigma_x^{m_1+m_3}$ on control and target, which can also be done by updating the Pauli frame.}
\label{fig:cnotcircuit}
\end{figure}

\begin{figure*}
\centering
\def\svgwidth{\linewidth}
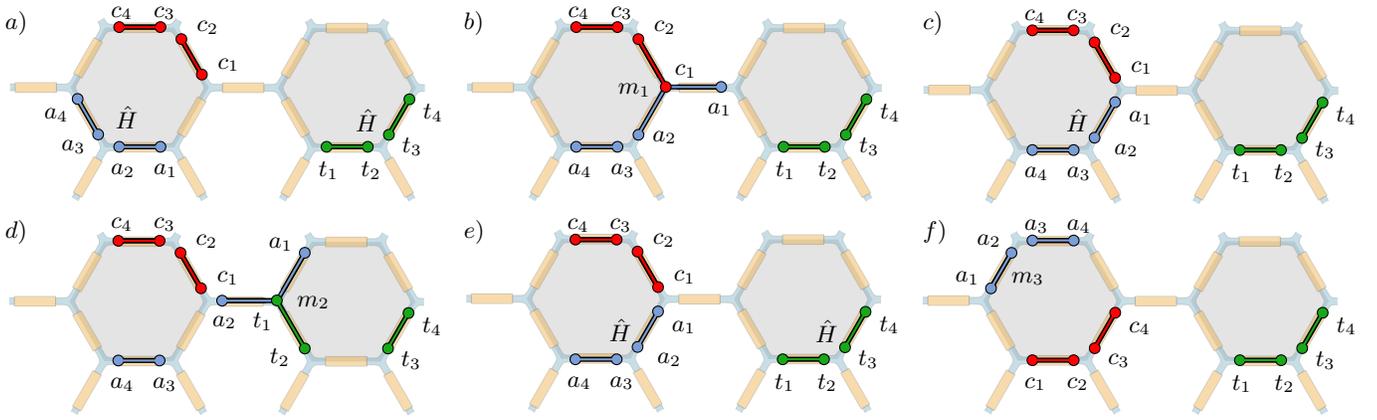
\caption{Protocol for a CNOT between two adjacent hexagonal cell qubits using the quantum circuit in Fig.~\ref{fig:cnotcircuit}. In the cell occupied by the control qubit (red), an ancilla (blue) is initialized in the $\ket{0}$ state and moved to the double T-junction of the cell. $a)$ The ancilla and target (green) are rotated via a Hadamard gate. $b)$ The first two Majoranas of the control $c_1$ and $c_2$ and ancilla $a_1$ and $a_2$ are moved onto a connected island and the four-Majorana fermion parity $-a_1a_2c_1c_2$ is measured, corresponding to a two-qubit parity measurement $\sigma_z \otimes \sigma_z$ with outcome $m_1$. $c)$ The ancilla is moved back to the double T-junction for another $H$ gate. $d)$ The ancilla and target parity $m_2$ is measured via a four-Majorana parity measurement in the right cell. $e)$ An $H$ gate is applied to the ancilla and target qubits in their respective double T-junctions. $f)$ Finally, all qubits return to their initial positions and the ancilla qubit is measured by measuring the two-Majorana fermion parity $ia_1 a_2$ with outcome $m_3$.}
\label{fig:hexcnot}
\end{figure*}

\begin{figure}[b]
\begin{tabular}{ccc}
\scalebox{1}{
\Qcircuit @C=1em @R=1.7em {
    \lstick{\ket{c}} & \ctrl{4} & \qw \\
     \\
    \lstick{\ket{t_1}} & \targ & \qw \\
    \lstick{\ket{t_2}} & \targ & \qw \\
    \lstick{\ket{t_3}} & \targ & \qw \\
    \raisebox{1em}{\vdots} & \raisebox{1em}{\vdots} & \raisebox{1em}{\vdots}
    }}
&
\raisebox{-5em}{$= \quad \quad$}
&
\scalebox{1}{
 \Qcircuit @C=1em @R=.7em {
    \lstick{\ket{c}} & \multigate{1}{P} & \qw & \qw & \qw & \gate{\sigma_z^{m_2}} & \qw \\
    \lstick{\ket{+}} & \ghost{P} &\gate{H} & \multigate{3}{P} & \gate{H} & \meter & \\
    \lstick{\ket{t_1}} &  \qw & \gate{H} & \ghost{P} & \gate{H} & \gate{\sigma_x^{m_1+m_3}} & \qw \\
    \lstick{\ket{t_2}} &  \qw & \gate{H} & \ghost{P} & \gate{H} & \gate{\sigma_x^{m_1+m_3}} & \qw \\
    \lstick{\ket{t_3}} &  \qw & \gate{H} & \ghost{P} & \gate{H} & \gate{\sigma_x^{m_1+m_3}} & \qw \\
     \vdots & \vdots & \vdots & \vdots & \vdots & \vdots & \vdots
    }}
    \put (-141,7) {$m_1$}
    \put (-93,-13) {$m_2$}
    \put (-18,-22) {$m_3$}
\end{tabular}
\caption{Quantum circuit for a multi-target CNOT gate using parity measurements and an ancilla qubit initialized in the $\ket{+}$ state. First, the parity operator $\sigma_z \otimes \sigma_z$ of control and ancilla is measured, with outcome $m_1$. Next, the parity operator $\sigma_x \otimes \sigma_x^{\otimes n}$ of ancilla and $n$ targets is measured, with outcome $m_2$. Finally, the ancilla is measured in the computational basis $\sigma_z$ with outcome $m_3$. The three outcomes determine the final correctional operation $\sigma_z^{m_2} \otimes (\sigma_x^{m_1+m_3})^{\otimes n}$ on control and targets, which can also be done by updating the Pauli frame.}
\label{fig:multicnot}
\end{figure}
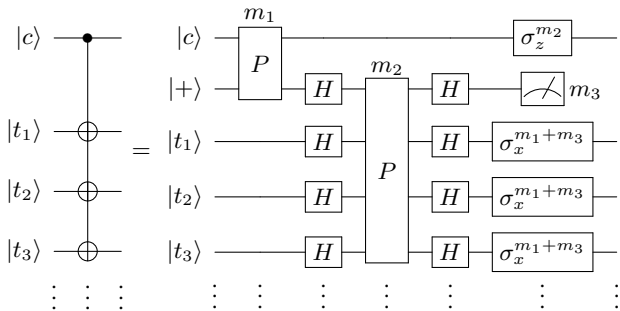

Braiding of Majoranas does not allow for a CNOT gate. However, qubit parity measurements and single-qubit Clifford gates can be used to construct a CNOT gate using an ancilla qubit~\cite{Zilberberg2008}. Consider the quantum circuit shown in Fig.~\ref{fig:cnotcircuit}. The action of a CNOT gate is to flip the target qubit $\ket{t}$ if the control qubit $\ket{c}$ is in the $\ket{1}$ state, and apply the identity if it is in the $\ket{0}$ state. Using an ancilla qubit initialized in the state $\ket{+} = (\ket{0}+\ket{1})/\sqrt{2}$ , a CNOT gate can be implemented by a series of qubit parity measurements and some corrective operations. In the first step of the quantum circuit, the two-qubit parity operator $\sigma_z \otimes \sigma_z$ between the control and  ancilla qubit is measured, yielding a measurement outcome $m_1 = 0$ for even and $m_1 = 1$ for odd parity. Next, the rotated parity operator $\sigma_x \otimes \sigma_x$ between ancilla and target qubit is measured with outcome $m_2$, which is equivalent to a $\sigma_z \otimes \sigma_z$ measurement with basis-rotating Hadamard gates applied before and after the measurement. Finally, the ancilla qubit is measured in the computational ($\sigma_z$) basis with outcome $m_3$. The three measurement outcomes are used to determine the correctional operation on the control and target qubit $\sigma_z^{m_2} \otimes \sigma_x^{m_1+m_3}$. This procedure can be seen as a topological version of the 
non-local CNOT gates considered in Ref.~\cite{NonLocalGates}.

Since the correctional operation consists of Pauli gates,
and Pauli gates can be commuted past Clifford gates generating only other Pauli gates, it is not necessary to physically perform the actual gate corresponding to the correction. Therefore, as long as the gate circuit consists only of Clifford operations,
Pauli gates only need to be tracked by a classical computer by updating the so-called Pauli frame, using a procedure known as 
Pauli tracking~\cite{PauliTracking}. 
This is strictly speaking no longer the case when $T$ gates are involved, since $\sigma_x T = T^\dagger \sigma_x$. In this case, gate synthesis at later steps needs to replace $T$ by $T^\dagger$ when commuting $\sigma_x$ past a $T$ gate.

This parity measurement-based protocol for a CNOT gate can be readily implemented in hexagonal cell qubits, see Fig.~\ref{fig:hexcnot}. First, the ancilla qubit is initialized in the hexagonal cell occupied by the control qubit. After the application of Hadamard gates on the ancilla and target qubit $(a)$, the two-qubit parity operator $\sigma_z \otimes \sigma_z$ of control and ancilla qubit is measured by moving the first two Majoranas of each qubit onto three connected superconducting islands $(b)$. Since the total fermion parity of the connected islands $(ic_1c_2)(ia_1a_2)$ is precisely the qubit parity operator, the measurement of the four-Majorana fermion parity yields the two-qubit parity. After another $H$ gate $(c)$, the same parity measurement is repeated for ancilla and target qubit $(d)$. After the final set of $H$ gates, all Majoranas are returned to their initial positions and the ancilla qubit is read out by measuring the two-Majorana parity $ia_1a_2$. The ancilla qubit may be discarded after the protocol. This concludes the protocol for a CNOT gate between adjacent hexagonal cell qubits. In Appendix~\ref{app:cnot}, we demonstrate that this scheme can also be used for CNOT gates between arbitrary hexagonal cell qubits. Moreover, we show that multiple CNOT gates applied in a transversal fashion can be applied simultaneously. 

This parity measurement-based protocol for a CNOT gate can be extended to multi-target CNOT gates. A multi-target CNOT gate corresponds to the application of $n$ CNOT gates with one control qubit $\ket{c}$ and $n$ different target qubits $\ket{t_i}$. Such multi-target CNOT gates are part of magic state distillation protocols, which are used for the implementation of a robust logical $T$ gate. Using the protocol in Fig.~\ref{fig:cnotcircuit}, a multi-target CNOT with $n$ targets would require $n$ ancilla qubits and $3n$ parity measurements. A faster alternative uses only one ancilla qubit and a parity measurement involving the ancilla and all $n$ target qubits, see Fig.~\ref{fig:multicnot}. This multi-target CNOT protocol replaces $3n$ measurements for $n$ CNOTs by just three measurements for an $n$-qubit multi-target CNOT. A proof of the circuit identity in Fig.~\ref{fig:multicnot} is given in Appendix~\ref{app:multicnot}. The application of this (transversal) multi-target CNOT gate to distillation protocols in topological superconductor networks is discussed in Sec.~\ref{sec:logqubits}.

\subsection{$T$ gates and stabilizer measurements}

So far, we have only shown the implementation of the non-universal set of Clifford gates in topological superconductor networks. In fact, by virtue of the Gottesman-Knill theorem, Clifford quantum computers are no more powerful than classical computers~\cite{Gottesman1999}. Unfortunately, the $T$ gate, which completes the universal gate set, cannot be done using a combination of braiding of Majoranas and parity measurements. An unprotected, error-prone $T$ gate can be achieved by splitting the degeneracy of the parity states on the island hosting the first two Majoranas, such that the energy splitting between $\ket{0}$ and $\ket{1}$ is $\Delta E$. After a time $\tau = \pi/4 \cdot \hbar / \Delta E$, the dynamic phase accumulated by time evolution will correspond to the $T$ gate, and the degeneracy is restored again.
In contrast to the Clifford gates, this protocol requires fine-tuning of the device control parameters and does not protect the $T$ gate against errors. There exist more sophisticated protocols for physical $T$ gates in Majorana-based setups~\cite{Karzig2015,Barkeshli2015}, but for our purposes, any implementation of physical $T$ gates is sufficient, as these gates can be used to implement $T$ gates with arbitrary precision using the magic state distillation procedure outlined in Sec.~\ref{sec:logqubits}.

\begin{figure}
\centering
\def\svgwidth{\linewidth}
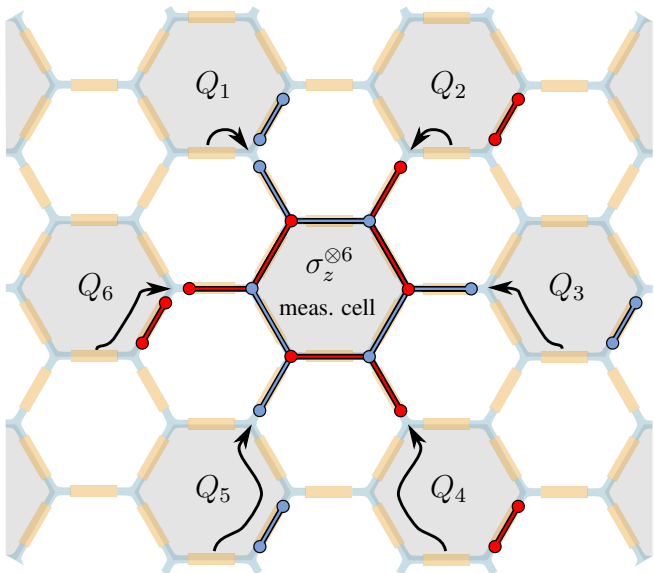
\caption{Six-qubit parity measurement in a triangular lattice of hexagonal cell qubits. Six hexagonal cell qubits $Q_1$ - $Q_6$, are arranged around an empty cell that is used for the measurement of the parity operator $\sigma_z^{\otimes 6}$. For clarity, the Majoranas of each qubit are colored red and blue in an alternating fashion. The first two Majoranas of each qubit are moved to this cell, such that 12 connected superconductors host 12 Majoranas. The total 12-Majorana parity of this island is precisely the six-qubit parity operator.}
\label{fig:hexstabilizer}
\end{figure}

In preparation for error correction using color codes, we also demonstrate the measurement of six-qubit parity operators $\sigma_z^{\otimes 6}$ without the need for ancilla qubits. Consider the six hexagonal cell qubits in Fig.~\ref{fig:hexstabilizer} arranged around an empty hexagonal cell. If the first two Majoranas of each surrounding qubit are moved onto 12 connected islands, the total parity of this island will be the 12-Majorana operator $\prod_{j=1}^6 i \gamma_{j,1}\gamma_{j,2}$, which is precisely the six-qubit parity operator $\sigma_z^{\otimes 6}$. This allows for the direct readout of the parity, circumventing the usual procedure~\cite{Fowler2012} involving an ancilla qubit and six CNOTs between the ancilla and each qubit. The measurement of such $n$-qubit parity operators is required for quantum error correction, where they are the stabilizers of the code.

In summary, we have shown that topological superconductor networks which allow for the movement of Majoranas, $2n$-Majorana parity measurements, and tuning of the energy splitting between parity states constitute universal quantum computers. In particular, triangular lattices of hexagonal cell qubits feature topologically protected Clifford gates and a $T$ gate requiring fine-tuning. Furthermore, $n$-qubit parity operators $\sigma_z^{\otimes n}$ can be measured without the need for ancilla qubits, and multi-target CNOT gates require only three parity measurements regardless of the number of target qubits.

\begin{figure*}
\centering
\def\svgwidth{\linewidth}
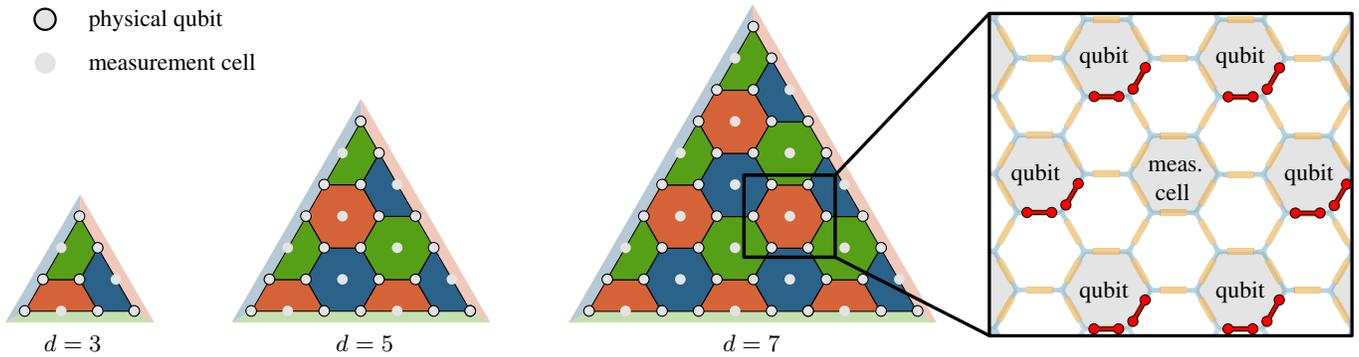
\caption{First three topological triangular color codes with code distances 3, 5 and 7 (where the smallest one is equivalent to the Steane code~\cite{Steane1996}). These 6.6.6 color codes are defined on a hexagonal lattice, where each vertex is a physical qubit and each face is an $X$-type and a $Z$-type stabilizer involving the surrounding qubits. Physical errors on a qubit affect the three different-colored stabilizers and edges surrounding the qubit. In the triangular lattice of hexagonal cell qubits, the empty cell in the center of each face can be used for stabilizer measurement.}
\label{fig:colorcode1}
\end{figure*}

\section{Topological software: \\ diamond color codes}
\label{sec:logqubits}

Unless the topological hardware is perfect, qubit errors will occur after a certain number of gate operations. These errors change the outcome of the quantum computation, and therefore need to be actively corrected. In \textit{quantum error correction}, multiple physical qubits are used to encode a single error-resilient logical qubit. In so-called \textit{stabilizer codes}~\cite{Gottesman1997,TerhalRMP}, the logical qubit is encoded in the degenerate ground state space of a Hamiltonian
\begin{equation}
	H_S = - \sum\limits_i \mathcal{O}_i \, , \quad [\mathcal{O}_i,\mathcal{O}_j] = 0 \, .
\end{equation}
Here, 
$\mathcal{O}_i$ are operators with eigenvalues $\pm 1$, which are called \textit{stabilizers} and are products of Pauli operators.
 Since all stabilizers commute, the ground state space is spanned by the simultaneous $+1$-eigenstates of all stabilizers, under the condition that the operator $-\mathbbm{1}$ is not part of the stabilizer group. Logical information can be stored in this degenerate ground state space, also referred to as code space. For the logical qubits discussed in this work, the ground state space is doubly degenerate, where the eigenstates define the logical qubit states $\ket{0_L}$ and $\ket{1_L}$. Note that the Hamiltonian $H_S$ does not necessarily describe the physical system used for quantum computation. Instead, $H_S$ merely defines the code space, into which the physical system is projected by measuring all stabilizer operators.

Errors occurring on physical qubits will change the eigenvalue of certain stabilizers.
The so-called \textit{code distance} is the minimum number of qubits that need to be affected by errors in order to change the logical subspace, i.e., map $\ket{0_L}$ onto $\ket{1_L}$ and vice versa. In order to prevent this from happening, all stabilizer operators are measured periodically before physical errors can affect the encoded information. These measurements reveal the so-called \textit{error syndrome}, which is a list of all stabilizer measurement outcomes $\pm 1$. 
This information is used to correct the errors that have occurred.
The practical problem that has to be 
overcome is that only the syndrome is available, while the actual errors are unknown.
Moreover, different error configurations can lead to the same error syndrome. 
The classical algorithm finding a suitable error configuration belonging to a given syndrome is called a
\textit{decoder}~\cite{Decoder1,Decoder2,Decoder3,Decoder4,Sarvepalli2011,Marks2017,Wang2009,Bombin2012}. 

\begin{figure*}
\centering
\def\svgwidth{\linewidth}
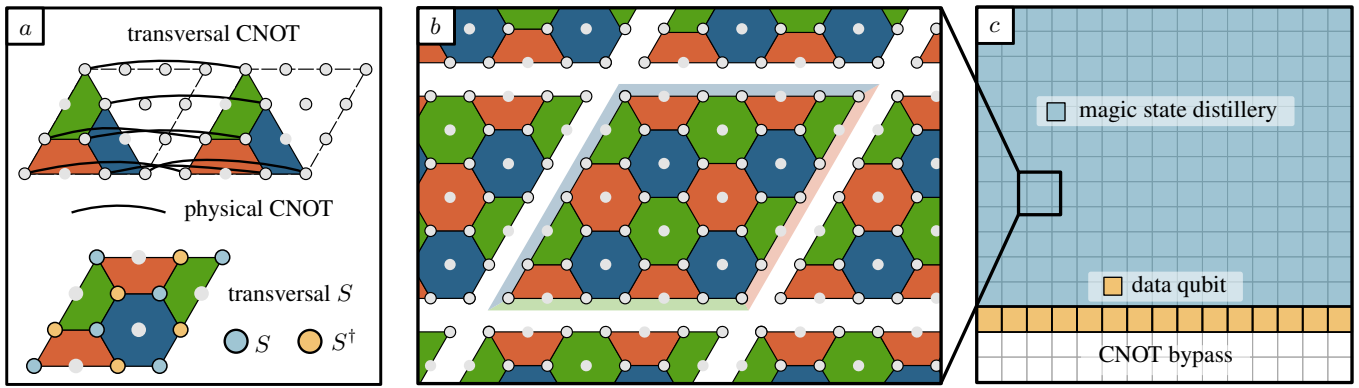
\caption{$a)$ Color codes feature transversal Clifford gates. While the logical Hadamard gate is simply $H_L = H^{\otimes n}$, the logical $S$ gate $S_L$ is a mixture of physical $S$ and $S^\dagger$ gates. Using a bicoloration of the physical qubits, such that the sublattice involving the corner qubits is blue and the other one is orange, $S_L$ requires physical $S$ gates on blue qubits and $S^\dagger$ gate on orange qubits. The logical CNOT gate corresponds to $n$ physical CNOTs between pairs of qubits from two triangles. $b)$ Since this requires the movement of ancilla qubits from one triangle to the other, this leaves some unused space in-between that can be used for the diamond color codes. These form a square lattice of logical qubits. $c)$~Universal fault-tolerant quantum computation can be achieved on a line of data qubits with a magic state distillery and a CNOT bypass.}
\label{fig:colorcode2}
\end{figure*}

Typically, quantum error-correcting codes operate in \textit{code cycles}. In every code cycle, logical operations are performed, 
the syndrome is read out by making use of stabilizer measurements 
and the errors on physical qubits are actively corrected. But even logical qubits only have a finite survival time, as quantum error-correcting codes merely replace a physical error rate by a (preferably lower) logical error rate. The minimum number of physical qubits that need to be affected by errors within a code cycle, such that the errors are no longer correctable, scales with the code distance. There are two prescriptions how a higher-distance code can be obtained from a low-distance code: code concatenation~\cite{Gottesman1997} and topological codes. Code concatenation has the drawback that it requires the measurement of increasingly non-local stabilizer operators with increasing code distance. In contrast, the stabilizers of topological codes remain spatially local as the code distance is increased. Moreover, in topological codes, the encoded logical quantum 
information is  protected from local perturbations because virtual transitions require an
order in perturbation proportional to the system size. In the case of surface and color codes, errors generate and propagate anyons~--~excitations of the system with nontrivial braiding statistics~--~that are manifested in a changed stabilizer measurement outcome. This implies that for surface and color codes defined on a lattice, anyons need to propagate through the entire lattice in order to affect the logical subspace, i.e., errors need to form along a nontrivial line through the lattice.
The locality of stabilizers and high error-resilience are the two key advantages that distinguish topological from non-topological codes.

In fault-tolerant quantum computing, it is desirable to perform all gate operations on the level of encoded logical qubits without the need to decode them back to error-prone physical qubits~\cite{Preskill1998}. However, the physical operations that constitute a logical gate $U_L$ are typically entirely different to the known physical gates $U$. An exception are so-called \textit{transversal gates}, which for our purposes are logical gates that are precisely the application of the corresponding physical gate (or its Hermitian conjugate) on each qubit, i.e., $U_L = U^{(\dagger)\otimes n}$. 
This has the advantage that errors due to faulty implementations of single physical gates do not spread to other physical qubits. Moreover, transversal gates directly employ physical gates to implement logical gates, enabling us to carry over the topological protection of physical gates to the level of logical gates. However, the Eastin-Knill theorem states that no code can have a set of transversal gates which is also a universal gate set~\cite{Eastin2008}.

One family of topological codes with transversal gates are topological color codes~\cite{Bombin2006}. Their set of transversal gates are the Clifford gates. Since this set coincides with the set of topologically protected operations of topological superconductor networks, color codes are a natural fit to Majorana-based hardware. In comparison to the closely related~\cite{Kubica2015} surface codes, which cannot implement braiding transversally, color codes also feature a higher \textit{error threshold}. The error threshold is the maximum physical error rate below which logical errors are suppressed by increasing the code size, allowing for quantum computation of arbitrary duration. We note that in circuit models where Clifford gates are error-prone and stabilizers are measured using ancilla qubits and CNOT gates, surface codes indeed feature a higher threshold than color codes~\cite{Campbell2016}. However, in the limit of topological hardware where Clifford operations have a vanishing error-rate and stabilizer readout does not require ancilla qubits, color codes outperform surface codes even in the presence of measurement errors during syndrome readout~\cite{Andrist2011,Andrist2016}. In addition, since the Clifford gates are transversal for color codes, their implementation only requires one code cycle. This reduces time overhead compared to surface codes, where their implementation requires multiple code cycles~\cite{Fowler2012}.

\subsection{Triangular and diamond color codes}

Color codes are stabilizer codes that are defined on lattices with 3-colorable faces. Physical qubits sit on the vertices and the stabilizers are operators 
acting on all qubits  surrounding a face. Figure~\ref{fig:colorcode1} shows a family of color codes that is defined on a hexagonal lattice of physical qubits, namely the triangular 6.6.6 color codes. Here, all stabilizers involve either four or six qubits. There are two stabilizers per face $f$, an $X$-type stabilizer $\mathcal{O}_X = \bigotimes_{i \in f}\sigma_{x}$ and a $Z$-type stabilizer $\mathcal{O}_Z = \bigotimes_{i \in f}\sigma_{z}$. Thus, the logical qubits in the color code are encoded in the ground state space of the Hamiltonian
\begin{equation}
	H_{\rm{color~code}} = - \sum\limits_{\rm faces} \mathcal{O}_X- \sum\limits_{\rm faces} \mathcal{O}_Z \, .
\end{equation}
To initialize a color code qubit in the logical $\ket{0_L}$ state, all physical qubits are initialized in the $\ket{0}$ state, the stabilizers are measured, and the errors corrected.

Every physical qubit is part of up to three different-colored $X$-type and $Z$-type stabilizers. At the boundaries, qubits are only part of one or two stabilizers, but if one assigns colors to the boundaries (see Fig.\ \ref{fig:colorcode1}), every qubit is part of three different-colored stabilizers \textit{or} boundaries. A $\sigma_z$-type Pauli error on a physical qubit will flip the three surrounding red, green and blue $X$-type stabilizers. Conversely, a $\sigma_x$-type error will flip three $Z$-type stabilizers. In the language of topological codes, flipped stabilizers with eigenvalue $-1$ host an anyon. Thus, errors generate and propagate strings with red, green and blue anyons at their endpoints. Each edge can absorb anyons of its respective color. A logical error occurs, when physical errors propagate a red, a green and a blue anyon to the red, green and blue edges respectively. Thus, a logical $(\sigma_z)_L$ operator is given by any string of physical $\sigma_z$ operators that propagates anyons in this way. In particular, physical $\sigma_z$ operators on all physical qubits sitting on any one of the three edges propagate anyons accordingly, and therefore correspond to logical $(\sigma_z)_L$ operators. Similarly, logical $(\sigma_x)_L$ operators correspond to strings of physical $\sigma_x$ operators.

\begin{figure*}
\centering
\def\svgwidth{\linewidth}
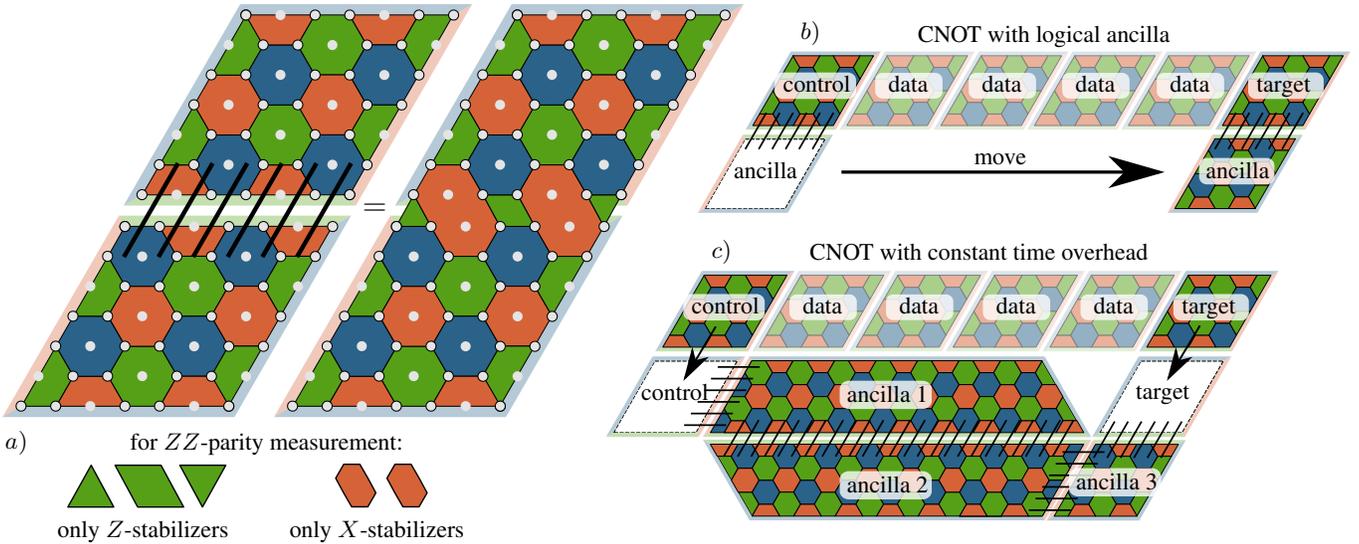
\caption{$a)$ Fault-tolerant $ZZ$-parity ($XX$-parity) measurement between two diamond color code qubits by lattice surgery~\cite{Landahl2014}, denoted by black lines crossing the neighboring boundaries. First, new three- and four-qubit green stabilizers are introduced, and new eight-qubit stabilizers are obtained by merging red plaquettes along the boundary. These stabilizers are measured along with all other stabilizers in order to obtain the $ZZ$-parity ($XX$-parity), where the new green boundary stabilizers are \textit{only} measured in the $Z$-basis ($X$-basis) and the red eight-qubit stabilizers only in the $X$-basis ($Z$-basis). (All stabilizers are explicitly shown in Fig. \ref{fig:surgeryappendix} in Appendix \ref{app:latticeSurgery}.) The product of the green boundary stabilizers is precisely the two-qubit parity. Finally, the stabilizers are returned to their initial configuration before the lattice surgery.
$b)$ Protocol for a fault-tolerant CNOT using lattice surgery. A logical ancilla is initialized in the $\ket{+}$ state. The $ZZ$-parity between control and ancilla is measured, and the ancilla is moved through the CNOT bypass to the target. Finally, the $XX$-parity between ancilla and target is measured, and the ancilla is read out. The length of this protocol scales linearly with the distance between control and target. $c)$ Lattice surgery-based CNOT protocol with constant time overhead. The $ZZ$-parities between control and three ancilla qubits in the $\ket{+}$ state are measured simultaneously using the three lattice surgeries indicated in the figure. Next, the $XX$-parity between ancilla 3 and target is measured. Finally, ancilla 3 is read out in the $Z$-basis, while ancillas 1 and 2 are measured in the $X$-basis. In the presence of measurement errors during syndrome readout, this protocol scales logarithmically with the distance between control and target.
}
\label{fig:surgerycnot}
\end{figure*}

Each code cycle consists of three steps. First, logical operations are performed on the encoded qubits. Next, the error syndrome is extracted by measuring all stabilizers. The syndrome is then given to the decoder. Finally, the corrections proposed by the decoder are applied.
Note that it is not necessary to physically correct the errors, as they can be handled classically by Pauli tracking~\cite{PauliTracking}, under the assumptions discussed in Sec.\ \ref{sec:physqubits}.

In the triangular lattice formed by hexagonal cell qubits, the cell in the center of each stabilizer is not occupied by a physical qubit. Instead, these cells can be used for stabilizer measurements, as shown in Fig.~\ref{fig:hexstabilizer} for $Z$-type stabilizers. $X$-type stabilizers can be measured by applying a Hadamard gate to all qubits before and after the measurement. Color codes fall into the class of CSS codes~\cite{Calderbank1996,Steane1996}, i.e., all stabilizers are products of only $\sigma_z$ operators or only $\sigma_x$ operators. CSS codes have a transversal implementation of the CNOT gate, where the logical CNOT gate corresponds to the application of physical CNOTs between all corresponding physical qubits of two codes, see Fig.~\ref{fig:colorcode2}a.
Moreover, color codes are \textit{strong} CSS codes, because the support of $Z$-type and $X$-type stabilizers coincides. This implies that Hadamard gates are transversal, and the logical Hadamard gate $H_L = H^{\otimes n}$ maps stabilizer states onto other stabilizer states. In general, this is not true for the application of physical $S$ gates on all qubits. Therefore, the transversal $S_L$ gate requires greater care, as some physical $S$ gates need to be replaced by $S^\dagger$ gates. One general prescription is to bicolor the vertices of the color code graph, such that neighboring qubits have different colors. In Fig.~\ref{fig:colorcode2}a, we color the sublattice containing the corner qubits blue, and the other sublattice orange. The logical $S_L$ gate then corresponds to physical $S$ gates on blue qubits and physical $S^\dagger$ gates on orange qubits~\cite{Kubica2014}.

All physical operations required for single-qubit transversal gates can be applied simultaneously, since they only require braiding within each hexagonal cell qubit. As we show in Appendix \ref{app:cnot}, also for transversal CNOTs, all physical CNOTs can be performed simultaneously in a hexagonal cell qubit geometry. However, this requires the triangles encoding the control and target qubit to be oriented the same way. Thus, the densest packing of triangular color codes along one line is not practical. Instead, we choose to extend the upward-pointing triangular codes into the unused space on their right, forming diamonds, as shown in Fig.~\ref{fig:colorcode2}b. Since all stabilizers of one type and color can be measured simultaneously, this happens at no increase in space or time overhead. Moreover, our Monte Carlo simulation in Sec.~\ref{sec:feasibility} shows that diamond codes even feature a lower logical error rate compared to triangular codes with the same code distance. 
Note that when extending triangles to diamonds, only one of the edges becomes longer compared to the triangular code. As the code distance is given by the length of the \textit{shortest} edge, the extension to diamond color codes lowers the logical error rate despite leaving the code distance unchanged.

Universal fault-tolerant quantum computation with logical diamond color code qubits requires the implementation of a universal gate set $\{H_L,S_L,\mathrm{CNOT}_L,T_L \}$. The first two gates are implemented directly in a  transversal fashion. The ${\rm CNOT}_L$ gate requires special care. Even though it can be done transversally, CNOTs in hexagonal cell qubits use physical ancilla qubits, which are not protected against noise.
In the remainder of this section, we show that a one-dimensional arrangement of data qubits with a magic state distillery above and a CNOT bypass below (see Fig.~\ref{fig:colorcode2}c) implements the remaining two logical gates in a fault-tolerant fashion. A magic state distillery is an array of qubits used for magic state distillation, whereas data qubits are qubits used for quantum computation but not for distillation. We present protocols that use the CNOT bypass to implement a fault-tolerant ${\rm CNOT}_L$ gate with an overhead that scales with neither the code distance nor the distance between the control and target qubits.
Furthermore, we demonstrate how the magic state distillery can be used to produce and store magic states, which allow for a fault-tolerant implementation of the $T_L$ gate.

\subsection{Logical CNOT gates}

We present three protocols for logical CNOT gates between data qubits. 
In the first protocol, the control qubit is moved to the target qubit through the CNOT bypass, and the CNOT gate is performed transversally, see Fig.~\ref{fig:colorcode2}a. However, the ancilla qubits in this protocol are physical qubits and therefore susceptible to errors. Moreover, since measurements are part of the CNOT protocol, a physical CNOT gate may introduce additional errors if measurements are not perfect. Both factors increase the noise level for logical CNOT gates. The noise level can be decreased by substituting physical ancilla qubits by a logical ancilla qubit.

An implementation of the circuit in Fig.~\ref{fig:cnotcircuit} with logical qubits requires parity measurements between logical qubits. Since the logical $\sigma_z$-operator is a nontrivial string of physical $\sigma_z$-operators through the code, the two-qubit $ZZ$-parity operator of two distance $d$ codes is a product of at least $2d$ $\sigma_z$-operators. One method of fault-tolerantly and projectively measuring the two-qubit parities of logical qubits is called lattice surgery~\cite{Landahl2014}. Here, new stabilizers are temporarily introduced on the boundary between two logical qubits. In Fig.~\ref{fig:surgerycnot}a, we show a lattice surgery protocol along the green boundaries of two diamond color code qubits, although any two boundaries can be used regardless of color as long as they have equal lengths. In this protocol, the red four-term $X$-stabilizers at the boundaries are merged to form eight-term stabilizers. The corresponding $Z$-stabilizers remain unchanged (see Fig.\ \ref{fig:surgeryappendix}
in Appendix \ref{app:latticeSurgery}). 
Green three- and four-term stabilizers are introduced which commute with all other stabilizers and involve each boundary qubit exactly once. Therefore, these stabilizers are only measured in the $Z$-basis, as the product of all green boundary stabilizers is precisely the $ZZ$-parity. If these stabilizers are measured along with the other stabilizers, qubit errors can be corrected and the parity measurement is fault-tolerant. The error due to faulty measurements can be reduced by repeating sufficiently many rounds of syndrome extraction. After the product of the green boundary stabilizers is determined~--~and therefore the two-qubit parity~--~the stabilizers are reverted to the initial configuration. Similarly, the $XX$-parity can be obtained by swapping $X$- and $Z$-stabilizers in the aforementioned protocol. We stress that the lattice surgery protocol projectively measures the logical two-qubit parity without revealing any additional information, as we discuss in greater detail in Appendix \ref{app:latticeSurgery}.

Thus, the second protocol is a CNOT with a logical ancilla shown in Fig.~\ref{fig:surgerycnot}b. A logical $\ket{+}$-ancilla is initialized in the CNOT bypass next to the control qubit. The $ZZ$-parity between ancilla and control is measured by lattice surgery, and the ancilla is moved to the target qubit. Finally, the $XX$-parity between ancilla and target is measured, and the ancilla is measured in $\sigma_z$-basis. Although this protocol yields a logical CNOT gate with arbitrary precision, it still has one major drawback: The protocol length increases linearly with the distance between the control and target qubit.

This can be alleviated by using two additional ancilla qubits with long edges, which replaces the movement of the ancilla qubit by a number of simultaneous stabilizer measurements at the long edge.
The third protocol is a CNOT with constant time overhead (see Fig.~\ref{fig:surgerycnot}c). Three $\ket{+}$-ancillas are arranged such that ancillas 1 and 2 both have a short and a long edge, and cover the entire distance between control and target qubit. Using lattice surgery, the $ZZ$-parities between control and ancilla 1, between ancillas 1 and 2, and ancillas 2 and 3 can be measured simultaneously. This is equivalent to measuring the two-qubit parities between the control qubit and each of the ancilla qubits. Therefore, ancilla 3 can be directly used as the CNOT ancilla. Its $XX$-parity with the target qubit is measured, and it is read out in $\sigma_z$-basis. Ancillas 1 and 2 cannot be discarded right away, as they are still entangled with the control qubit. They can be disentangled by measuring the ancillas in the $\sigma_x$-basis with measurements outcomes $m_1$ and $m_2$, and applying a $\sigma_z^{m_1+m_2}$-correction to the control qubit. An explanation of the quantum circuit corresponding to this protocol is found in Appendix \ref{app:constanttimecnot}.

In the absence of measurement errors, this protocol has a constant time overhead. This is no longer true, if syndrome measurements are faulty. Such a measurement can be 
described by a perfect measurement, followed by the identity map with probability $p$ and a flipped
outcome with probability $1-p$. 
Since more boundary stabilizers are involved in the parity measurement involving ancillas 1 and 2, they need to be measured more often to achieve the same accuracy as the other parity measurements. However, because the measurement error probability decreases exponentially with each repetition, whereas the number of boundary stabilizers only increases linearly with the distance between control and target, the time overhead of this CNOT only scales logarithmically with the control-target distance.

We have presented three protocols for logical CNOT gates. The transversal protocol between nearest-neighbors is fast, but has a fixed accuracy and a time overhead that scales linearly with the control-target separation. The second protocol uses a logical ancilla, and can therefore achieve arbitrary accuracy, but is slower than the first protocol as it requires multiple code cycles. The third protocol eliminates the time overhead, or replaces it by a time overhead that scales favorably as the logarithm of the distance between control and target. By adding rows to the CNOT bypass, multiple spatially intertwined CNOT gates can be performed simultaneously. Note that due to the overhead in quantum wires in a hexagonal cell qubit, logical diamond color code qubits do not block each other's paths when moving, as they can be moved through one another, similar to how ancilla qubits can be moved past other qubits in the transversal CNOT protocols of Figs.~\ref{fig:hexcnot} and \ref{fig:transversalcnot}.

\subsection{Magic state distillation}

\begin{figure}[b]
\begin{tabular}{ccc}
\raisebox{-0.9em}{\scalebox{1}{
\Qcircuit @C=1em @R=.7em {
    \lstick{\ket{\psi}} & \gate{T} & \qw 
    }}}
&
\raisebox{-1.15em}{$= \quad \quad$}
&
\scalebox{1}{
 \Qcircuit @C=1em @R=.7em {
    \lstick{\ket{\psi}} & \ctrl{1} & \gate{S^{m_z}} & \qw  \\
    \lstick{\ket{m}} & \targ & \meter & & &
    }}
    \put (-26,-21) {$m_z$}
\end{tabular}
\caption{State injection algorithm: A CNOT between a qubit $\ket{\psi}$ and a magic state $\ket{m} =  (\ket{0} + e^{i\pi/4}\ket{1})/\sqrt{2}$, followed by a measurement of $\ket{m}$ with outcome $m_z$ and a correctional $S^{m_z}$ gate is equivalent to a $T$ gate on the qubit.}
\label{fig:stateinjection}
\end{figure}
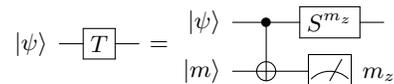

The only gate remaining for a universal fault-tolerant quantum computer is the logical $T_L$ gate. We point out that even if the physical hardware had a topologically protected physical $T$ gate, there would be no way of directly using this for a $T_L$ gate as the $T$ gate cannot be transversal in a code with transversal Clifford gates due to the Eastin-Knill theorem~\cite{Eastin2008}, which states
that the ability of a quantum code to detect arbitrary errors on any single physical subsystem is incompatible with the existence of a universal, transversal encoded gate set for the code.
There exist \textit{code switching methods} that allow to switch the logical qubit from one code to another code with a different set of non-universal transversal gates. However, in order for this set to include the $T$ gate and for the stabilizers to still remain local, the qubits need to be arranged in \textit{three} dimensions instead of two~\cite{Pastawski2015}.

One possibility to implement a logical low-error $T$ gate using logical Clifford gates and a physical $T$ gate is magic state distillation. Consider the state injection circuit shown in Fig.~\ref{fig:stateinjection}, which is equivalent to a $T$ gate on the qubit $\ket{\psi}$. Using an ancilla magic state $\ket{m} = T\ket{+} =\left(\ket{0} + e^{i\pi/4}\ket{1}\right)/\sqrt{2}$, a CNOT between the qubit and one prepared in a magic state followed by the measurement of the magic state with outcome $m_z$ corresponds to a $T$ gate up to a correctional $S^{m_z}$ operation. Such a procedure
of effectively generating a quantum gate by making use of suitable quantum state resources is referred to as \textit{gate teleportation}.

In order for this state injection algorithm to yield a logical $T$ gate, the magic state $\ket{m}$ needs to be an encoded logical qubit. However, since the physical $T$ gate is not topologically protected and physical qubits are not protected against errors during the encoding process, we can only generate faulty magic states that are well-approximated by $ \ket{\widetilde{m}} = \left(\ket{0} + e^{i(\pi/4 + \varepsilon)}\ket{1}\right)/\sqrt{2}$, even though further errors are expected and allowed for.
 Magic state distillation is an algorithm deeply related to quantum error correction that generates low-error magic states using many faulty magic states $\ket{\widetilde{m}}$ with angle deviations $\varepsilon$ of up to $17.3 \%$~\cite{Bravyi2005}. Many such algorithms exists, such as a 15-to-1 protocol~\cite{Bravyi2005}, a 10-to-2 protocol~\cite{Meier2013}, or more generally for an integer $k$ a
 3$k$+8-to-$k$ protocol~\cite{Bravyi2012,BlockCodeStateDistillation}. These protocols only require (transversal) Clifford gates, in particular 
 multi-target CNOT gates. Combinations of these protocols~\cite{Bravyi2012} can be used to generate magic states~--~and therefore 
 effectively $T$ gates~--~with the desired precision.

\begin{figure}
\centering
\def\svgwidth{\linewidth}
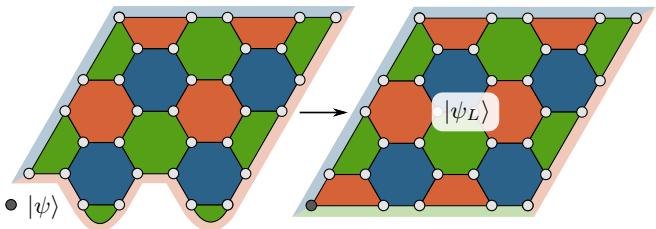
\caption{Code injection procedure which encodes an unknown physical state $\ket{\psi}$ (gray qubit) into a logical state $\ket{\psi_L}$. First, the stabilizer state in the left panel is prepared by measuring all shown stabilizers. Finally, we cease measuring the green stabilizers at the bottom boundary, and start measuring the red stabilizers.}
\label{fig:injection}
\end{figure}

Logical magic states can be encoded from physical magic states using a variant of
the code injection procedure described in Ref.~\cite{Landahl2014}.
In Fig.~\ref{fig:injection}, we depict this procedure for a diamond color code. A detailed explanation of the presented protocol is given in Appendix \ref{app:stateinjection}. The protocol can correct errors on any pair of physical qubits, but certain errors with support on three qubits cause the injection of a faulty state, regardless of the code distance of the used diamond code. This further substantiates the need for magic state distillation.

In principle, the multi-target CNOTs in the distillation protocols can be done using many iterations of the logical CNOT gates that we discussed previously. However, for logical CNOTs between data qubits, we focused on these operations having a low error-rate. Since distillation protocols are only performed once, and afterwards magic state qubits are merely stored until their use, the priority of their multi-target CNOTs should be speed over accuracy of individual gates, such that magic states can be distilled fast.

\begin{figure}
\centering
\def\svgwidth{\linewidth}
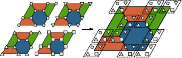
\caption{Inflation protocol for transversal multi-target CNOT gates with four logical qubits. This protocol rearranges the physical qubits such that the qubits involved in transversal multi-target CNOT gates are now close to each other, i.e., every first physical qubit of each of the four logical qubits, every second, etc. This protocol can be used for the multi-target CNOTs required for magic state distillation, e.g., using inflation of 15 qubits arranged on a $4 \times 4$ grid for 15-to-1 distillation.}
\label{fig:inflation}
\end{figure}

Majorana-based qubits offer the possibility of a fast multi-target CNOT gate using the protocol in Fig.~\ref{fig:multicnot}. Even though this gate is transversal for color code qubits, the parity measurements involve physical qubits that are spatially separated~--~i.e., every first physical qubit of each involved logical qubit, every second physical qubit, and so on. One method to bring them closer together is by rearranging the physical qubits using the inflation protocol shown in Fig.~\ref{fig:inflation} for the example of four logical qubits arranged on a $2\times 2$ grid. The protocol effectively rearranges the physical qubits of four logical qubits, such that they form blocks of four physical qubits that are part of multi-target CNOT gates. The analogous protocol with 15 qubits arranged on a $4\times 4$ grid can be used for the transversal multi-target CNOTs required for 15-to-1 distillation. After sufficiently many rounds of magic state distillation, the magic state is ready for state injection via a CNOT gate using any of the protocols outlined in the previous subsection.

Clifford gates and magic state distillation operate independently from each other. That is, during the application of Clifford gates on the data qubits in the quantum computation, magic states can be distilled in parallel and stored for later use in the magic state distillery. Magic states can even be prepared offline and stored for future quantum computations. Since magic state distillation is the part of the quantum computation that requires the greatest effort, magic states are resource states for quantum computation. With pre-distilled magic states, any quantum computation reduces to the application of (constant time overhead) logical Clifford gates.

In conclusion, we have constructed logical diamond-shaped color code qubits with transversal Clifford gates. Arranged on a line with a CNOT bypass and a magic state distillery, they feature a robust $T$ gate and a CNOT gate with constant time overhead. The single-qubit Clifford gates are topologically protected due to the protection of the topological superconductor network. We note that apart from transversal CNOTs and fast multi-target CNOTs, the remaining protocols make no use of the diamond shape. In fact, if for data qubits one abandons the fast transversal CNOT protocol, each diamond-shaped data qubit can be replaced by two triangular color code qubits with a straightforward generalization of the lattice surgery protocols. This reduces the spatial overhead for data qubits by a factor of two, but also slightly increases the logical error rate. The same is not true for magic state distillery qubits, as the inflation protocol for fast distillation still benefits from diamond color codes.

\section{Physical architecture: \\ Majorana Cooper pair boxes}
\label{sec:hardware}

In the previous sections, we have demonstrated that we can construct a fault-tolerant universal topological quantum computer on the basis of a topological superconductor network. Our construction requires that 
Majoranas can be moved, their parities measured, and the degeneracy of their parity states lifted. In this section, we review how this can be achieved using Majorana Cooper pair boxes following the scheme suggested in Ref.\ \cite{Aasen2016} (see also~\cite{Vijay2015,Hassler2011,VanHeck2012,Hyart2013}).  While here, we follow Ref.\ \cite{Aasen2016}, other implementations of Majorana qubits can also be combined with a color code as discussed in this paper, as long as these architectures are capable of the three required operations. For instance, this scheme can in principle also be realized in Majorana box qubits~\cite{Plugge2016a} and related setups~\cite{Karzig2016}, where Majoranas are not moved directly by coupling neighboring islands, but via braiding by measurement~\cite{Bonderson2008,Vijay2016a}.

\begin{figure}[t]
\centering
\def\svgwidth{\linewidth}
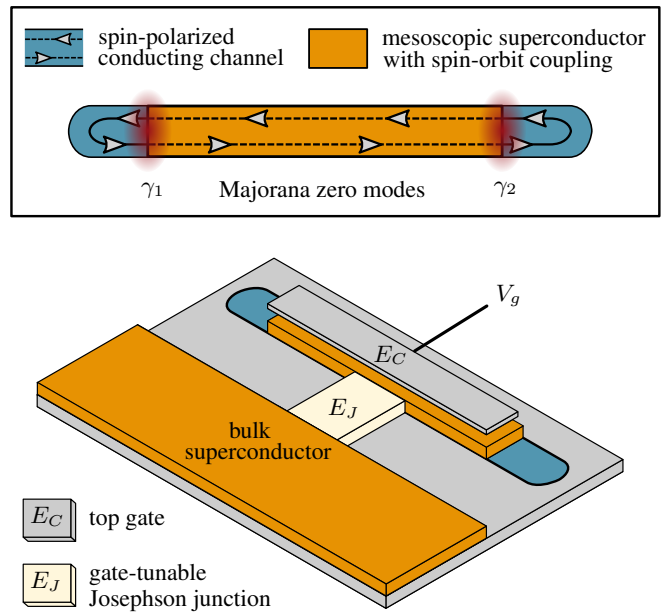
\caption{A Majorana Cooper pair box as a basic building block of the topological hardware. Top: A pair of Majorana zero modes $\gamma_1$ and $\gamma_2$ at the ends of a p-wave superconductor can be effectively obtained by depositing an s-wave superconductor with strong spin-orbit coupling on top of a material with a single spin-polarized conducting channel, such as a semiconducting nanowire in a magnetic field, a quantum anomalous Hall insulator or a 2DEG in a strong magnetic field. Bottom: A Majorana Cooper box requires the addition of charging energy $E_C$ and Josephson energy $E_J$ on the mesoscopic superconducting island. A top gate which is capacitively coupled to the superconducting island imposes a certain total charge on the island governed by the gate voltage $V_g$ and the (fixed) charging energy. Furthermore, a bulk superconductor is Josephson-coupled to the mesoscopic island through a gate-tunable Josephson junction, which tunes the Josephson energy and imposes a certain phase on the island.}
\label{fig:setup}
\end{figure}

Pairs of Majorana zero modes can emerge at the ends of one-dimensional spinless p-wave superconductors~\cite{Kitaev2001}. Even though there are candidates for p-wave superconductors such as $\mathrm{Sr}_2\mathrm{RuO}_4$~\cite{Mackenzie2003}, ordinary superconductors exhibit s-wave pairing. To effectively obtain the required p-wave pairing from s-wave pairing~--~and thereby Majorana zero modes~--~three essential ingredients are required (see Fig.\ \ref{fig:setup}): an s-wave superconductor, spin-orbit coupling, and one-dimensional spin-polarized conducting channels~\cite{Fu2008,Oreg2010,Lutchyn2010,Alicea2012,Leijnse2012,Beenakker2013}. Experiments have focused on realizing this by using nanowires~\cite{Mourik2012,Das2012,Albrecht2016a,Deng2016} or by appropriate patterning of two-dimensional electron gases~\cite{Shabani2016,Suominen2017}, but in principle, this could also be achieved in edge states of quantum Hall, quantum spin Hall, or quantum anomalous Hall systems~\cite{Fu2008,Amet2016,Lee2016,Chang2013,Liu2016}.  

Unlike in ordinary s-wave superconductors, where the minimal excitation energy is given by the pairing gap $\Delta$, the Majoranas have zero excitation energy. Each pair of Majoranas combines into a complex fermion that can be empty or occupied. Unpaired electrons can occupy these fermionic states at zero energy cost. When the island has one Majorana at each end, there is one complex zero-energy fermion. The occupation of this energy level is associated with the fermion parity of the mesoscopic island, i.e., the level is unoccupied for even and occupied for odd fermion parity. These statements hold true when the Majorana wire is proximity coupled to a grounded s-wave superconductor. If the superconductor is floating, the combined system of wire and proximity coupled superconductor has a finite charging energy which will in general lift the degeneracy between the even and odd parity states~\cite{Fu2010,Beri2012,Altland2014}. 

\begin{figure*}[t]
\centering
\def\svgwidth{\linewidth}
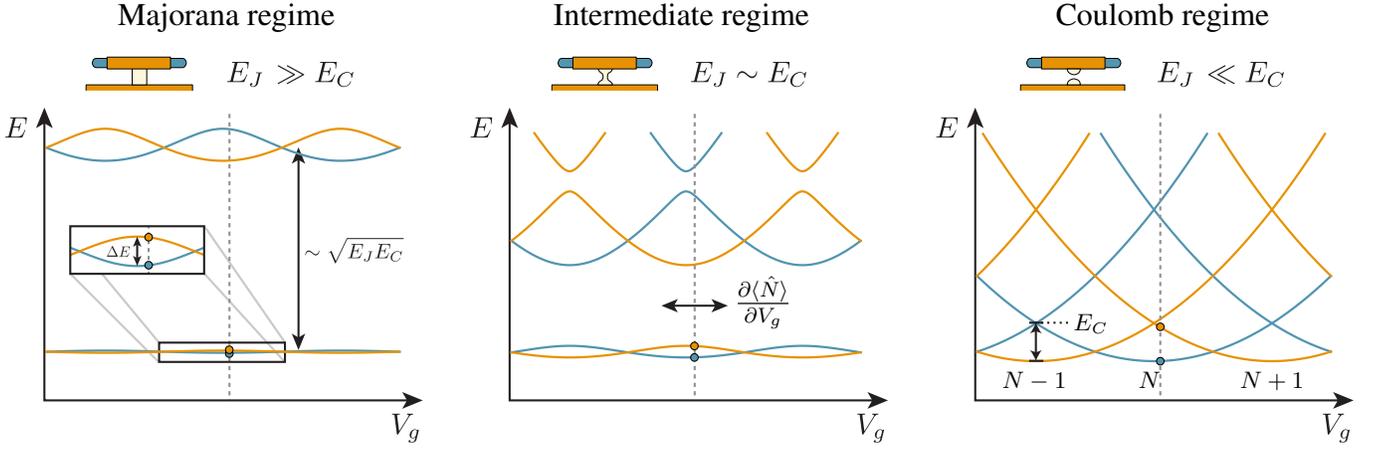
\caption{Parity-to-charge conversion in the Majorana Cooper pair box, as described in Ref.~\cite{Aasen2016}, and energy levels of the Majorana Cooper pair box $H_C + H_J$ as a function of gate voltage $V_g$ and for different ratios $E_J/E_C$. In the Majorana regime $E_J \gg E_C$, charging energy is negligible and the spectrum is insensitive to $V_g$. The ground state is given by nearly-degenerate states of opposite parity (blue and orange), where the maximum separation $\Delta E$ vanishes exponentially in $E_J/E_C$, whereas the distance to the first excited states increases with $\sqrt{E_J E_C}$. As $E_J$ is decreased by decreasing the coupling to the bulk superconductor, the ground state degeneracy is lifted in the intermediate regime. Here, varying the gate voltage distinguishes the parity states through their differential capacitance $C=\partial \langle \hat{N} \rangle/\partial V_g$, which is larger for the orange parity than for the blue parity. Finally, in the Coulomb regime $E_J \ll E_C$, the spectrum is given by parabolas with well-defined charge number $N$. If $V_g$ is tuned to a minimum of a charge parabola, the two lowest-energy parity states are separated by $E_C$, which can be used to impose a certain parity on the island. The intermediate regime can also be understood from the emergence of avoided crossings between charge states of equal parity as $E_J$ is increased.}
\label{fig:dispersion}
\end{figure*}

A powerful scheme to manipulate Majorana zero modes exploits Majorana Cooper pair boxes, see Fig.\ \ref{fig:setup}~\cite{Aasen2016}. A gated wire coated by a superconducting island is coupled to a bulk superconductor through a tunable Josephson junction. Opening the Josephson junction effectively grounds the island which will then support a Majorana degeneracy. This degeneracy will be progressively lifted by Coulomb charging effects as the Josephson coupling is reduced. 

The low-energy Hamiltonian of the Majorana Cooper pair box~\cite{Aasen2016} is given by the sum $H=H_C+H_J$ of a charging term 
\begin{equation}
	H_C = E_C \left(\hat{N} - N_0\right)^2
\end{equation}
with charging energy $E_C$, and a Josephson term
\begin{equation}
	H_J = - E_J \cos \hat{\varphi} \, ,
\end{equation}
with Josephson energy $E_J$. Here, $\hat{N}$ is the operator that counts the electrons on the island and $N_0 = e V_g /(2E_C$) is the background charge controlled by the gate voltage $V_g$ applied to the capacitively coupled gate. The operator $\hat{\varphi}$ is the phase of the superconducting island, obeying the commutation relation $[\hat{\varphi},\hat{N}] = 2i$. Even and odd parity states obey periodic and antiperiodic boundary conditions when writing the wave function in the phase representation~\cite{Fu2010}. 

\begin{figure}[b]
\centering
\def\svgwidth{\linewidth}
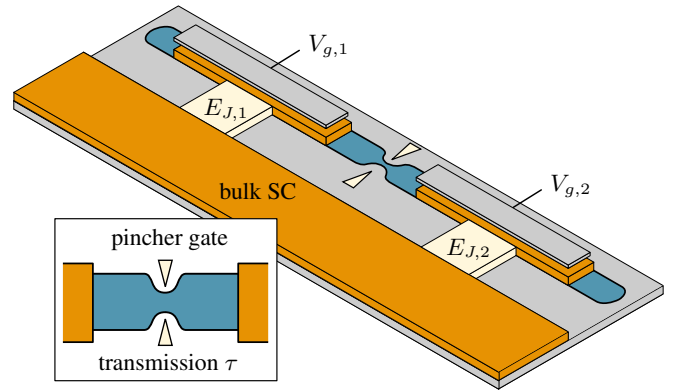
\caption{Two Majorana Cooper pair boxes connected to the same bulk superconductor with Josephson energies $E_{J,1}$ and $E_{J,2}$, and top gate voltages $V_{g,1}$ and $V_{g,2}$ respectively. The islands are connected through the spin-polarized conducting channel, in which the inter-island transmission probability $\tau$ can be tuned by a pincher gate.}
\label{fig:twoislands}
\end{figure}

Figure~\ref{fig:dispersion} shows the spectrum of $H$ in three characteristic regimes~\cite{Aasen2016}. In the \textit{Majorana regime} $E_C\ll E_J$, the phase $\hat{\varphi}$ is fixed by the bulk superconductor and the spectrum is almost $V_g$-independent. In this regime, there are two nearly degenerate ground states whose splitting $\Delta E$ is exponentially small in $E_J/E_C$. These ground states are separated from excited states by an energy $\sim \sqrt{E_J E_C}$. In the opposite \textit{Coulomb regime} $E_C\gg E_J$, the eigenstates are well-defined charge states. The two lowest charge states with even and odd parity are split for all values of $V_g$, except at the charge degeneracy points where $N_0$ is half integer. Depending on whether $N_0$ is closer to an even or odd integer, the ground state has either even or odd fermion parity. Thus, one can impose a desired fermion parity on the state of the Majorana Cooper pair box by tuning it to the Coulomb regime and relaxation to the ground state. The \textit{intermediate regime} with $E_C\sim E_J$ can be understood starting from the Majorana regime as the result of Coulomb charging lifting the ground-state degeneracy or from the Coulomb regime as the result of forming avoided crossings between states of equal fermion parity by Cooper pair tunneling in and out of the island.

Using these three regimes of the Majorana Cooper pair box, all operations required for color code quantum computing with a topological superconductor network can be implemented. In a network, islands hosting Majoranas that encode a qubit are tuned to the Majorana regime, such that the parity states~--~and therefore the encoded qubits~--~are degenerate. All other (empty) islands are tuned to the Coulomb regime. The remainder of this section is devoted to showing how to use these two regimes to move Majoranas through the network and how to employ the intermediate regime for parity measurements~\cite{Aasen2016}. This is complemented by degeneracy splitting which is straightforwardly implemented by decreasing $E_J$ on an island.

\subsection{Moving Majoranas}

\begin{figure}
\centering
\def\svgwidth{0.93\linewidth}
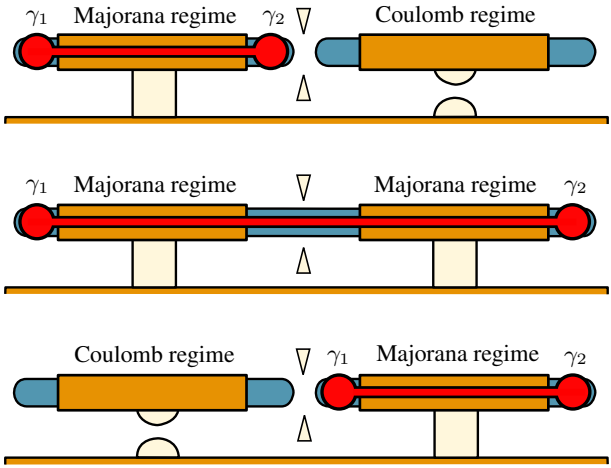
\caption{Two-step protocol for moving Majoranas $\gamma_1$ and $\gamma_2$ from the left island, initially tuned to the Majorana regime, to the right island, initially tuned to the Coulomb regime with even parity. First, the two islands are coupled by increasing the transmission to $\tau=1$ and tuning the right island to the Majorana regime, shuttling $\gamma_2$ to the right island. The two islands now form a single connected superconducting island with Majoranas $\gamma_1$ and $\gamma_2$. In order to move $\gamma_1$ to the right island, the transmission is reduced back to $\tau = 0$, and the left island is tuned to the Coulomb regime.}
\label{fig:twoislandmove}
\end{figure}

Neighboring Majorana Cooper pair boxes with individually controllable gate voltage and Josephson energy are connected via tunnel coupling, see Fig.~\ref{fig:twoislands}. The inter-island transmission probability $\tau$ can be controlled by a pincher gate located between the islands. Following Ref.~\cite{Aasen2016}, this junction can be used to move Majoranas between islands. Starting with two decoupled islands ($\tau = 0$) in the Majorana (left island) and Coulomb (right island) regime (see Fig.~\ref{fig:twoislandmove}), $\gamma_2$ is moved to the right island by increasing the inter-island coupling, and then tuning the right island to the Majorana regime by increasing its Josephson coupling to the bulk superconductor. This places the system into an eigenstate of the \textit{total} parity $i\gamma_1\gamma_2$ of both islands. One should ensure that at the beginning of the protocol, the right island is initialized into the even parity sector by tuning $V_g$ accordingly, so that moving the Majorana will not flip the parity state. Finally, $\gamma_1$ can be moved to the right island by tuning the left island into the Coulomb regime, and then decoupling the two islands.

\begin{figure}
\centering
\def\svgwidth{\linewidth}
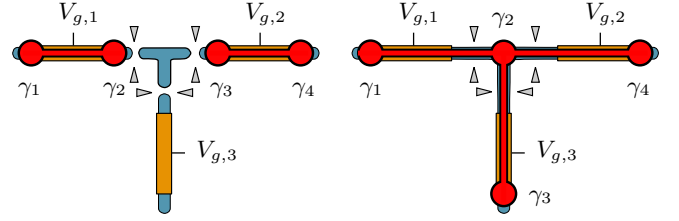
\caption{T-junction geometry consisting of three mesoscopic superconducting islands coupled through a three-terminal junction involving three pincher gates. Left: If all three pincher gates are closed, the islands are decoupled and host a pair of Majoranas each. In dispersive readout, tuning $V_{g,1}$ and $V_{g,2}$ is used to measure the parities $i\gamma_1\gamma_2$ and $i\gamma_3\gamma_4$ respectively. Right: Opening the right and bottom pincher gate while tuning the bottom island into the Majorana regime moves $\gamma_3$ to the bottom island. Subsequently opening the left pincher gate connects all three islands, where $\gamma_2$ is shared by all islands in the three-terminal junction.  Connecting any of the top gates to a resonant circuit allows for dispersive readout of the total parity $-\gamma_1\gamma_2\gamma_3\gamma_4$.
}
\label{fig:threeislands}
\end{figure}

Majorana Cooper pair boxes arranged in a T-junction geometry with three pincher gates between three islands (see Fig.\ \ref{fig:threeislands}) form the basic building block of our proposed network and implement all required  moving operations. Opening any pair of pincher gates couples the respective islands. For instance, opening the right and bottom pincher gates in the left configuration of Fig.~\ref{fig:threeislands} moves $\gamma_3$ from the right island to the bottom island. Opening the remaining pincher gate connects the left island to the other two superconductors, such that $\gamma_2$ becomes a Majorana shared by all three islands.

\subsection{Fermion parity measurements}

The set of required operations is completed by measurements of the fermion parity of $2n$ Majoranas. The fermion parity $i \gamma_1 \gamma_2$ of an island can be measured by tuning to the Coulomb regime and measuring the charge on the island (parity-to-charge conversion). In an alternative scheme, the superconducting island is tuned into the intermediate regime $E_J \sim E_C$ with the gate voltage set such that $N_0$ is, say, an even number. Then, the even parity state is at the minimum of a charge parabola, while the odd parity state sits at an avoided crossing between two charge parabolas. Consequently, the charge on the island is insensitive against variations of the gate voltage in the even parity state, but susceptible in the odd state, i.e., the two parity states differ in the differential capacitance
\begin{equation}
	C = \frac{\partial \langle \hat{N} \rangle}{\partial V_g}.
\end{equation}
When incorporating the island into a resonant circuit, the resonant frequency depends on the differential capacitance. Thus, a measurement of the resonance frequency constitutes a parity measurement, referred to as dispersive readout~\cite{Persson2010,Stehlik2015,Petersson2010,Colless2013}. 

This dispersive readout scheme can be generalized to measuring the fermion parity of 2$n$ Majoranas. Moving the $2n$ Majoranas onto one connected superconducting island which is tuned away from the Majorana regime with suitably chosen gate voltage, the parity can be read out by incorporating this island into a resonant circuit and proceeding as before. An experimental limitation is set by the decrease of the charging energy with increasing island size. Typically, for nanowire networks, the charging energy decreases linearly with the system size. Therefore, the required precision in the gate voltage control increases linearly with the size of the island. This scheme can for instance be applied to measure the four-Majorana parity operator $-\gamma_1 \gamma_2 \gamma_3 \gamma_4$. In the right configuration of the T-junction in Fig.\ \ref{fig:threeislands}, all pincher gates are opened and the three islands form one connected mesoscopic superconductor hosting four Majorana zero modes. The spectrum of this effective Majorana Cooper pair box will be the same as in Fig.~\ref{fig:dispersion}, but with a correspondingly lower charging energy and states differing in \textit{total} parity $(i\gamma_1\gamma_2)(i\gamma_3\gamma_4)$. Thus, dispersive readout now measures this four-Majorana parity operator. 

In this manner, the $2n$-Majorana parity operator of an island hosting $2n$ Majoranas can be measured via dispersive readout by connecting any of the top gates to a resonant circuit, and measuring the quantum capacitance~\cite{Persson2010}. Since at least one top gate in each hexagonal cell needs to be connected to its own readout circuit, incorporating the readout hardware in a two-dimensional architecture is an experimental challenge. Similar to superconducting qubit platforms, few-qubit quantum computers may allow for an on-chip implementation of both qubit and readout hardware. However, as each gate in the two-dimensional platform needs to be addressed individually, large-scale quantum computing will require the gates to be contacted to the control hardware via the third dimension, i.e., out of plane. In fact, the integration of the control and measurement hardware in a three-dimensional architecture is the subject of current experimental efforts~\cite{Bejanin2016,Liu2017,Rosenberg2017} to scale up superconducting qubit platforms.

Alternative schemes that were proposed for the parity readout of Majorana-based qubits~\cite{Aasen2016,Plugge2016a,Karzig2016} include charge reflectometry, where a resonator is used to probe the island's energy levels as opposed to the differential capacitance, and charge sensing, which employs the Coulomb regime to read out the average charge on the island. The implementation and benchmarking of different parity readout techniques is the subject of ongoing experimental efforts.

Having discussed the implementation of the operations required of topological superconductor networks, we now turn to investigating error sources of Majorana Cooper pair boxes and how well they can be corrected by diamond color codes in the following section.

\section{Feasibility estimate}
\label{sec:feasibility}

The performance of diamond color code qubits in topological superconductor networks depends on the error sources of Majorana Cooper pair boxes. Even though the parity states are degenerate in the Majorana regime for $E_J \gg E_C$, a finite overlap between Majorana wavefunctions on one island will split the degeneracy. Still, this splitting is exponentially suppressed in the island size. Overlap between Majoranas of neighboring islands can also lead to errors, but the overlap is proportional to the controlled tunneling amplitude between neighboring islands and is thus also exponentially small.

An error that is not necessarily exponentially suppressed occurs when an outside electron tunnels onto an island. This process is called quasiparticle poisoning, which is presumably the dominant error source in Majorana-based qubits. In the following, we model poisoning on any of the two islands encoding a physical qubit by the application of one of the four Majorana operators. This does not only change the total parity sector of the qubit, but also leads to a logical Pauli error depending on the Majorana involved in the process. The change of the parity sector is inconsequential to the qubit, since in the encodings of both parity sectors in Eqs.\ \eqref{eqn:qubits1} and \eqref{eqn:qubits2}, the physical qubit operators are $\sigma_z = i\gamma_1 \gamma_2$ and $\sigma_x = i\gamma_2 \gamma_3$. Therefore, merely switching the parity sector leaves both the logical information and the logical braid operations $B_{1,2}$ and $B_{2,3}$ unchanged. However, $\gamma_1$ anticommutes with $\sigma_z$, $\gamma_2$ anticommutes with $\sigma_z$ and $\sigma_x$, and $\gamma_3$ anticommutes with $\sigma_x$. Therefore, poisoning of $\gamma_1$ leads to a $\sigma_x$-error, of $\gamma_2$ to a $\sigma_y$-error, of $\gamma_3$ to a $\sigma_z$-error, and of $\gamma_4$ to no error. We discuss this in further detail in Appendix \ref{app:poisoning}. Moreover, we discuss more general error sources that are not described by a single Majorana operator.

Since $\sigma_y$-errors correspond to both a $\sigma_x$ and a $\sigma_z$-error, the quasiparticle poisoning time defines a time-scale, on which $\sigma_x$-type and $\sigma_z$-type errors occur at equal rates. Current experiments suggest that the quasiparticle poisoning time of mesoscopic superconducting islands might be of the order of milliseconds~\cite{VanWoerkom2015,Higginbotham2015,Albrecht2016}, although we point out that these experiments were performed in a regime where the superconducting islands were not floating but connected to a pair of normal-metal leads. We note that even though the regime of equally likely $\sigma_x$- and $\sigma_z$-type errors is the one considered in the following discussion, this is actually the worst-case scenario for error correction.
If one error type is known to occur more often, these errors have been shown to be correctable with fewer resources~\cite{Robertson2017}, albeit by changing the code and therefore giving up on transversal gates. But even without abandoning color codes, a biased error source can be taken into account by measuring the corresponding syndrome type more frequently than the other, thereby reducing the code cycle duration and hence the error rate.

In non-topological architectures, random Pauli errors are usually not a realistic error model, since relaxation processes from excited states to ground states are not described by unitary operations. For topological hardware, on the other hand, there are no transitions between different parity states that would allow for relaxation from one qubit state to the other. Therefore, we believe that random Pauli errors should be a reasonable error model for topological physical qubits. With this error model, the physical error threshold for color codes is $\sim 11 \%$~\cite{Andrist2016,Marks2017}, where the physical error rate is the probability for a physical error on one physical qubit after one code cycle. For our physical hardware, this physical error rate is $p_{\rm phys} = 1 - e^{-\tau_c/\tau_p} \approx \tau_c/\tau_p$, where $\tau_p$ is given by the quasiparticle poisoning time and $\tau_c$ is the duration of a code cycle. As moving Majoranas can be done at nanosecond time-scales~\cite{Knapp2016} without introducing significant diabatic errors, the code cycle duration is mainly determined by the time required for parity measurement. Dispersive readout on superconducting qubits suggests that this can be done on microsecond time-scales~\cite{Persson2010,Stehlik2015} or faster. For quasiparticle poisoning times of the order of milliseconds, the physical error rate would be $p_{\rm phys} \approx 10^{-3}$, which is well below threshold.

In order to estimate the survival time of logical qubits and the performance gain of diamond color codes over triangular color codes, we use a Monte Carlo simulation of the quantum error-correcting code for the aforementioned error model using a lookup table decoder. We note that our decoder does not take correlations between $\sigma_x$- and $\sigma_z$-errors into account, which could further enhance the correction procedure with a suitable decoder. A detailed discussion of the simulation is found in Appendix~\ref{app:numerics}. In Fig.~\ref{fig:numerics}, we show the logical error rate as a function of physical error rate for the first three lowest-distance triangular color codes and first two diamond color codes. The simulation reproduces the error threshold of $\sim 11 \%$, and shows that the logical error rate of diamond color codes is indeed lower compared to triangular color code of the same code distance. Furthermore, we find that already for the $d=5$ diamond color code of Fig.~\ref{fig:colorcode2}b with $p_{\rm phys} = 10^{-3}$, the survival time of logical qubits is approximately $35000$ code cycles until the probability for a logical error reaches $1\%$. In order to determine the survival time for larger code distances, a more efficient decoder needs to be used, such as an iterative decoder~\cite{Sarvepalli2011} or a color clustering decoder~\cite{Marks2017}. Both slightly lower the error threshold to $7.8 \%$ and $9.75 \%$ respectively. Furthermore, a decoder may keep track of multiple rounds of syndrome extraction in order to take measurement errors into account. While it is not known how readout errors affect the logical error rate, their effect on the error threshold has been studied~\cite{Andrist2016}. As a concrete example, a 95\% readout fidelity lowers the error threshold from $\sim \! 11\%$ to 2.5\%. A numerical study of the corresponding logical error rate would help quantify the performance of color codes, but goes beyond the scope of this work.

Still, we may extrapolate our results to at least estimate the survival time for higher-distance codes. Details on this are found in Appendix~\ref{app:numerics}. The extrapolation suggests that for $p_{\rm phys} = 10^{-3}$, $\tau_c = 1 {\rm \mu s}$, and the more stringent requirement that the logical error probability stays below $10^{-6}$, the $d=19$ diamond color code has a survival time of several years, implying that with an overhead of roughly 500 physical qubits per logical qubit quantum computations may run for reasonably long durations. We note that diamond color code qubits are not resource-efficient in the number of physical qubits, but only a useful construction for transversal CNOTs and multi-target CNOTs, as discussed in Sec.\ \ref{sec:logqubits}. Since equal-distance triangular color codes do not have a substantially higher logical error rate, the number of physical qubits per logical qubit can be reduced by a factor of two, if data qubits are encoded using triangles instead of diamonds.

\begin{figure}
\centering
\def\svgwidth{\linewidth}
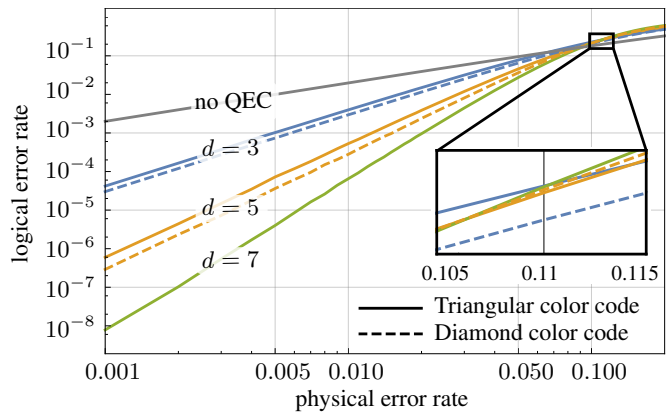
\caption{Logical error rate as a function of physical error rate obtained from Monte Carlo simulation with a lookup table decoder for triangular (solid line) and diamond (dashed line) color codes with code distances $d=3$ and $d=5$, and for the triangular color code with $d=7$. The sample size is between $10^7$ and $10^{10}$ trials for data point corresponding to high and low logical error rates respectively. The upper line show the logical error rate without quantum error-correction. The inset zooms into the crossover region around $p_{\rm phys}\sim 11\%$.}
\label{fig:numerics}
\end{figure}

\section{Conclusion}
\label{sec:conclusion}

In this work, we have studied the interplay of topological hardware and topological error-correcting software. Using topological superconductor networks, we have devised a scalable architecture for universal fault-tolerant topological quantum computation, which can be realized with voltage-controlled Majorana Cooper pair boxes as basic building blocks. The underlying physical qubits are hexagonal cell qubits, which allow for universal quantum computing with topologically protected Clifford gates, fast multi-target CNOT gates, and ancilla-free syndrome readout. For quantum error-correction, we employ topological color codes. Their set of transversal gates coincides with the topologically protected Clifford gates, which enables the logical gates to retain their topological protection due to the topological hardware. This makes color codes a natural fit to Majorana-based hardware, as they seamlessly combine topological hardware with topological software while still benefiting from the topological protection of both. Moreover, color codes also feature a reduced time-overhead for gate operations and a higher error threshold compared to surface codes, even in the presence of measurement errors during stabilizer readout. In a qubit arrangement consisting of a row of data qubits, a magic state distillery, and a CNOT bypass, logical single-qubit Clifford gates have a fast transversal implementation, CNOTs between any pair of data qubits have a constant time overhead,
and magic states can be distilled faster using transversal multi-target CNOT gates.
Our architecture is not restricted to implementations using Majorana Cooper pair boxes, but can be applied to any realization of a topological superconductor network, provided that Majoranas can be moved, that their parities can be measured and that some implementation of a physical $T$ gate is available.

Considering the particular geometry of a Majorana-based color code quantum computer presented in this work, i.e., hexagonal cell qubits and 6.6.6 diamond color codes, we make no claim of this geometry being optimal in terms of space and time overhead. Studies of different network layouts and color code schemes may reduce the overhead. Still, it is not clear how different code layouts and decoders affect the logical error rate. In particular, 4.8.8 color codes require fewer physical qubits compared to 6.6.6 codes with the same code distance. However, as we show in Appendix \ref{app:488comp}, they also feature a higher logical error rate, even though they have the same code distance and error threshold. 
Similarly, for the comparison of triangular and diamond codes, neither code distance nor error threshold are predictive figures of merit for logical error rates. We therefore encourage studies of the logical error rate of topological codes, in order to quantify the performance of codes beyond the already well-studied error thresholds and code distances.
Moreover, in order to further quantify the performance of a topological color code quantum computer, it would be interesting to estimate the number of code cycles required for actual computational tasks in an arrangement of data qubits and magic state distilleries.

On the hardware side, the past years have shown considerable experimental progress towards the realization of 
Majorana zero modes through the interplay of superconductivity, spin-orbit coupling and one-dimensional spin-polarized channels. 
This work is expected to provide further motivation for ongoing efforts to achieve braiding of Majoranas in these systems. 
On a more general note, aiming at merging ideas of both hardware- and software-based topological protection,
we hope that our work further stimulates research efforts bringing 
the fields of condensed matter physics and quantum information theory closer together.

\section*{Acknowledgments}

We thank F.\ Pastawski, S.\ Acero, A.\ Bauer, M.\ Filippone, M. Geier, M.\ Gluza, A.\ Haim, T.\ Karzig, Y.\ Oreg, D.\ Sabonis, Y.\ Vinkler-Aviv and other participants
of the CRC 183 for illuminating discussions. This work has been supported by the DFG (CRC 183),
the ERC (TAQ), the Templeton Foundation, and 
the EC (AQuS).

\appendix

\section{Protocol for a transversal CNOT gate}
\label{app:cnot}

\begin{figure*}
\centering
\def\svgwidth{\linewidth}
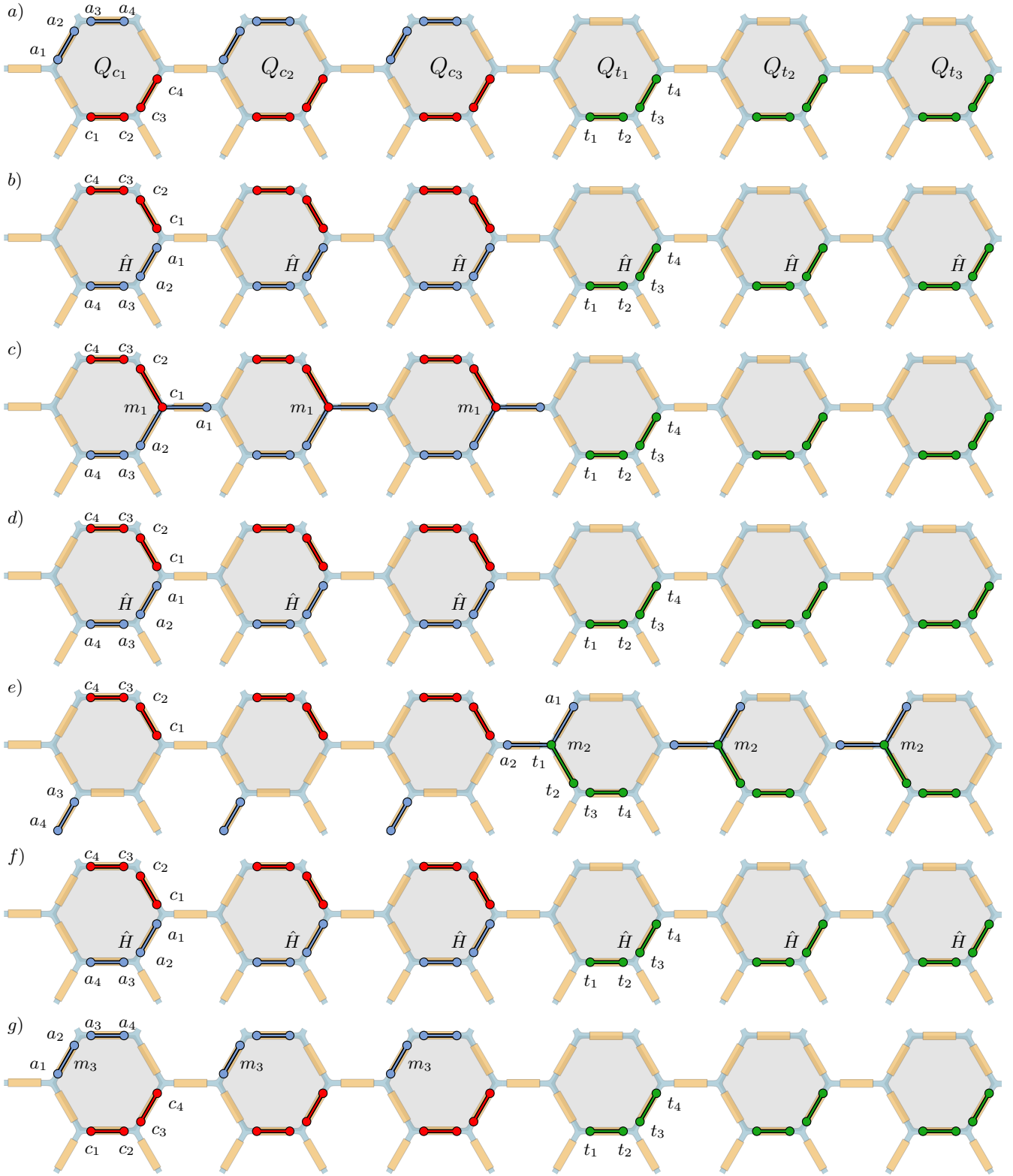
\caption{Protocol for three simultaneous CNOTs between three control qubits $Q_{c_1 - c_3}$ and three target qubits $Q_{t_1-t_3}$ using the quantum circuit in Fig.~\ref{fig:cnotcircuit}. In the cell occupied by the control qubits (red), ancillas (blue) are initialized in the $\ket{0}$ state $(a)$ and moved to the double T-junction of the cell for a Hadamard gate $(b)$. The first two Majoranas of the control and ancilla qubits are moved onto a connected island and the four-Majorana parity is measured $(c)$, corresponding to a two-qubit parity measurement with outcome $m_1$. The ancillas are moved back to the double T-junction for another $H$ gate $(d)$. The third and fourth Majoranas $a_3$ and $a_4$ of each ancilla qubit are moved into the lower leg of their hexagonal cell, such that the remaining ancilla Majoranas can move to the target qubit cells for a four-Majorana parity measurement $(e)$. The ancilla Majoranas are then moved back to the control cells for an $H$ gate  $(f)$. Finally, all qubits return to their initial positions $(g)$ and the ancilla qubits are measured by measuring the two-Majorana parity $m_3$.}
\label{fig:transversalcnot}
\end{figure*}

In Fig.~\ref{fig:transversalcnot}, we show that in hexagonal cell qubits, CNOT gates between control and target qubits arranged on a line can be performed simultaneously. 
This is a generalization of the protocol introduced in Fig.~\ref{fig:hexcnot} using the quantum circuit in Fig.~\ref{fig:cnotcircuit}. Since in diamond color codes, control and target qubits are also arranged on a line, all transversal CNOT gates in color codes can be performed simultaneously.

\section{Multi-target CNOT by parity measurement}
\label{app:multicnot}

Here, we show that a multi-target CNOT gate can be realized using Clifford gates and qubit parity measurements, see Fig.~\ref{fig:multicnot}. The multi-target CNOT operator
\begin{equation}
	{\rm CNOT}_n = \ket{0}\bra{0} \otimes \mathbbm{1}^{\otimes n} + \ket{1} \bra{1} \otimes \sigma_x^{\otimes n}
\end{equation}
flips all $n$ target qubits if the control qubit is in the $\ket{1}$ state. An $n$-qubit parity measurement with outcome $m=0$ for even and $m=1$ for odd parity is equivalent to an operation
\begin{equation}
	P_z = \frac{1}{2} \left[ \mathbbm{1}^{\otimes n} + (-1)^m \sigma_z^{\otimes n} \right] \, .
\end{equation}
Similarly, an $n$-qubit parity measurement in the $\sigma_x$-basis is
\begin{equation}
	P_x = H^{\otimes n}P_zH^{\otimes n} = \frac{1}{2} \left[ \mathbbm{1}^{\otimes n} + (-1)^m \sigma_x^{\otimes n} \right] \, .
\end{equation}

\begin{table}[t]
\bgroup
\def\arraystretch{1.5}
\begin{tabular}{c|c|c|c}
$m_1+m_3$ & $m_2$ & $U_{m_1+m_3,m_2}$ & Correction\\
\hline
0 & 0 & $\ket{0}\bra{0} \otimes \mathbbm{1}^{\otimes n} + \ket{1} \bra{1} \otimes \sigma_x^{\otimes n}$ & $\mathbbm{1} \otimes \mathbbm{1}^{\otimes n}$\\
0 & 1 & $\ket{0}\bra{0} \otimes \mathbbm{1}^{\otimes n} - \ket{1} \bra{1} \otimes \sigma_x^{\otimes n}$ & $\sigma_z \otimes \mathbbm{1}^{\otimes n}$\\
1 & 0 & $\ket{1}\bra{1} \otimes \mathbbm{1}^{\otimes n} + \ket{0} \bra{0} \otimes \sigma_x^{\otimes n}$ &  $\mathbbm{1} \otimes \sigma_x^{\otimes n}$\\
1 & 1 & $\ket{1}\bra{1} \otimes \mathbbm{1}^{\otimes n} - \ket{0} \bra{0} \otimes \sigma_x^{\otimes n}$ & $\sigma_z \otimes \sigma_x^{\otimes n}$
\end{tabular}
\egroup
\caption{Uncorrected gate $U_{m_1+m_3,m_2}$ and necessary correction based on measurement outcomes $m_1$, $m_2$ and $m_3$.}
\label{tab:multicnot}
\end{table}

Thus, the circuit in Fig.~\ref{fig:multicnot} in the basis $\ket{c}\otimes\ket{a}\otimes\ket{t}^{\otimes n}$, where $c$, $a$, and $t$ denote the control, ancilla, and the $n$ target qubits respectively, is
\begin{equation}\begin{split}
	U = & \left(\mathbbm{1} \otimes \frac{1}{2}\left[\mathbbm{1} + (-1)^{m_3} \sigma_z \right] \otimes \mathbbm{1}^{\otimes n}\right) \\
	\times &\left(\mathbbm{1} \otimes \frac{1}{2}\left[\mathbbm{1}^{\otimes n+1} + (-1)^{m_2} \sigma_x^{\otimes n+1} \right] \right) \\
	 \times &\left(\frac{1}{2}\left[\mathbbm{1}^{\otimes 2} + (-1)^{m_1} \sigma_z^{\otimes 2} \right] \otimes \mathbbm{1}^{\otimes n} \right) \\
	\times & \left(\mathbbm{1} \otimes \ket{+}\bra{+} \otimes \mathbbm{1}^{\otimes n} \right) \, ,
\end{split}\end{equation}
where the final correction is not yet applied. Tracing out the ancilla qubit yields 
\begin{equation}\begin{split}
	U = \quad &\frac{1}{2} \left( \mathbbm{1} + (-1)^{m_1+m_3} \sigma_z \right) \otimes \mathbbm{1}^{\otimes n} \\
	 + \, &\frac{1}{2} (-1)^{m_2} \left( \mathbbm{1} - (-1)^{m_1+m_3} \sigma_z \right) \otimes \sigma_x^{\otimes n} \, .
\end{split}\end{equation}
Depending on the measurement outcomes $m_1+m_3$ and $m_2$, there are four possible uncorrected operations $U_{m_1+m_3,m_2}$, see Tab.~\ref{tab:multicnot}. Thus, after the correction $\sigma_z^{m_2} \otimes \left(\sigma_x^{m_1+m_3}\right)^{\otimes n}$, the circuit in Fig.~\ref{fig:multicnot} precisely yields the multi-target CNOT gate ${\rm CNOT}_n$ using only three measurements, as opposed to $3n$ measurements for $n$ individual CNOTs.

\section{Details on lattice surgery}
\label{app:latticeSurgery}

\begin{figure}[t]
\centering
\def\svgwidth{\linewidth}
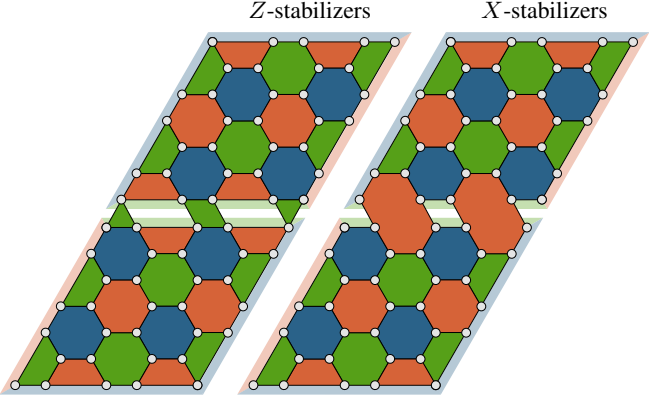
\caption{Stabilizers that are measured to obtain the $ZZ$-parity between two logical qubits using lattice surgery. In contrast to usual color code stabilizer measurements, lattice surgery requires measurements where the support of $X$- and $Z$-stabilizers does not coincide. The left panel shows the required $Z$-stabilizers, and the right panel the $X$-stabilizers, which differ along the shared boundary. To obtain the $XX$-parity between two qubits, one simply has to swap $Z \leftrightarrow X$ in the protocol.}
\label{fig:surgeryappendix}
\end{figure}

The stabilizers that are measured during lattice surgery~\cite{Landahl2014}  are shown in Fig.\ \ref{fig:surgeryappendix}. In the following, we describe the protocol for $ZZ$-parity measurement, but this can be straightforwardly used for $XX$-parity measurement by simply swapping $Z \leftrightarrow X$ in the protocol.
As one can verify
by direct inspection, in this $ZZ$-parity measurement between two distance $d$ codes, $\lceil\frac{d}{2}\rceil$ new $Z$-stabilizers are introduced and $2\lfloor\frac{d}{2}\rfloor$ $X$-stabilizers are replaced by $\lfloor\frac{d}{2}\rfloor$ octagon stabilizers. Since $d$ is always odd, exactly one new stabilizer is introduced. This reduces the number of logically encoded qubits by one, implying that in this process, one bit of information is measured. In the 
following, we would like to make the case that the measured bit is precisely the $ZZ$-parity.

First, notice that in the absence of errors, the extended octagon stabilizers will not detect anyons (i.e., have a measurement outcome $+1$), since they are products of pre-existing stabilizers. In this error-free setting, the only stabilizers that give a non-trivial measurement outcome are the $2\lfloor\frac{d}{2}\rfloor$ newly created $Z$-stabilizers. The product of these stabilizers is exactly the logical two-qubit parity $(\sigma_z\sigma_z)_L$. Another way to understand this 
process is in terms of anyons. If in the error-free setting, an anyon is detected on one of the new $Z$-stabilizers (depicted
as green in Fig.\ \ref{fig:surgeryappendix}), this means that the number of strings from the upper code terminating on said plaquette differs in parity from the number of strings coming from the lower code. Thus, the total parity of all newly created plaquettes measures exactly the difference of strings from the upper and lower code, which is precisely the two-qubit parity.

Lastly, in order to convince oneself of the fault tolerance of the above process, it is sufficient to check that strings corresponding to logical operations in the new settings still involve at least $d$ physical qubits. In Fig.\ \ref{fig:surgeryappendix}, this has to be fulfilled for logical $(\sigma_x)_L$, $(\sigma_z)_L$ and $(\sigma_x\sigma_x)_L$ operations, but not for $(\sigma_z\sigma_z)_L$, since this commutes with the parity measurement.

\section{Details of color code state injection}
\label{app:stateinjection}

Figure \ref{fig:injection} shows the protocol for the injection of a single qubit state $\ket{\psi}$ into a logical state $\ket{\psi_L}$ encoded in a diamond color code. This is a direct adaptation of the protocol in Ref.\ \cite{Landahl2014}, where this protocol was introduced for the specific case of
triangular 4.8.8 color codes. Here, we explain how this protocol achieves the state injection adapted to our situation.

The left panel of Fig.\ \ref{fig:injection} depicts the stabilizers measured before state injection. The number of stabilizers is exactly the same as the number of physical qubits and thus no logical qubits can be encoded. Since all stabilizers are measured and errors are corrected for, no anyonic excitations are present initially. The fact that there are only two boundaries implies that an even number of strings has to leave each boundary. 
This is a consequence of errors always creating pairs of anyons of the same color or triples of all three colors (and combinations thereof), and 
of the fact that boundaries can only host anyons of their respective color.

In the concrete example shown in Fig.\ \ref{fig:injection}, the $Z$-parity of all qubits along the blue boundary is even, $\sigma_z^{\otimes n} = +1$. The same holds for the $X$-parity and equivalent measurements along the red boundary. Importantly, this statement generalizes and holds for all color codes with two boundaries, regardless of code distance, geometry and tiling. 

To inject the state of the single physical qubit into the color code, the stabilizers shown in the right panel of Fig.\ \ref{fig:injection} are measured. Even if no errors on physical qubits occur, the new red plaquettes might still host anyons. Importantly, they are not corrected according to the most likely error configuration producing this syndrome; instead, they are moved over the red boundary. If errors occur, they will manifest themselves 
in the syndrome readout and can be corrected. 

The blue boundary after state injection differs only by the addition of the new physical qubit. Thus, measurements of the logical state along this boundary are given by the state of the new physical qubit alone. The way in which anyons on the new red plaquettes are corrected ensures that the same holds for all other measurements of logical operations as well. This proves that the protocol successfully injects the state of the single physical qubit $\ket{\psi}$ into the logical state $\ket{\psi_L}$ encoded in the color code.

\section{Constant time overhead CNOT}
\label{app:constanttimecnot}

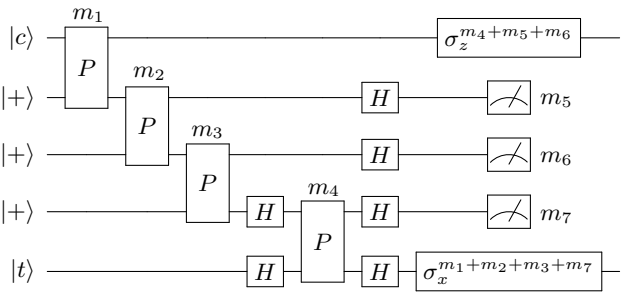
\begin{figure}[b]
\scalebox{1}{
 \Qcircuit @C=0.75em @R=1em {
    \lstick{\ket{c}} & \multigate{1}{P} & \qw & \qw & \qw & \qw & \qw & \gate{\sigma_z^{m_4+m_5+m_6}} & \qw \\
    \lstick{\ket{+}} & \ghost{P} &\multigate{1}{P} & \qw & \qw & \qw &\gate{H} & \meter & \\
    \lstick{\ket{+}} &  \qw & \ghost{P} & \multigate{1}{P} & \qw & \qw & \gate{H} & \meter & \\
    \lstick{\ket{+}} &  \qw & \qw & \ghost{P} & \gate{H} & \multigate{1}{P} & \gate{H} & \meter & \\
    \lstick{\ket{t}} &  \qw & \qw & \qw & \gate{H} & \ghost{P} & \gate{H} & \gate{\sigma_x^{m_1+m_2+m_3+m_7}} & \qw 
    }}
    \put (-207,7) {$m_1$}
    \put (-184,-15) {$m_2$}
    \put (-162,-37) {$m_3$}
    \put (-118,-59) {$m_4$}
    \put (-30,-25) {$m_5$}
    \put (-30,-47) {$m_6$}
    \put (-30,-69) {$m_7$}
\caption{Quantum circuit corresponding to the constant time overhead CNOT gate in Fig.\ \ref{fig:surgerycnot}c.}
\label{fig:surgerycnotcircuit}
\end{figure}

The quantum circuit in Fig.\ \ref{fig:surgerycnotcircuit} describes the constant time overhead CNOT protocol in Fig.\ \ref{fig:surgerycnot}c. The protocol involves a control qubit $\ket{c}$, a target qubit $\ket{t}$ and three $\ket{+}$-ancillas, where the third ancilla may be thought of as the ancilla that is part of the CNOT protocol of Fig.\ \ref{fig:cnotcircuit}. The $ZZ$-parity between this third ancilla and the control qubit is not measured directly, but as the sum of the first three parity measurements in the circuit $m_1+m_2+m_3$. The $XX$-parity between the third ancilla and the target is measured with outcome $m_4$ and the third ancilla is read out with outcome $m_7$. This is the reason for the $\sigma_x^{m_1+m_2+m_3+m_7}$ correctional operation on the target and the $\sigma_z^{m_4}$-correction on the control qubit.
However, these operations alone leave the first two ancilla qubits entangled with the control qubit in a state of the type $\ket{\psi} = \alpha \ket{0,0,0} + \beta \ket{1,1,1}$. In order to safely discard the two ancilla qubits without affecting the control qubit, they are measured in the $X$-basis with outcomes $m_5$ and $m_6$, leading to a \linebreak $\sigma_z^{m_5+m_6}$-correction on the control qubit.

\section{Quasiparticle poisoning}
\label{app:poisoning}

In the following, we discuss qubit errors due to quasiparticle poisoning. We find that merely changing the parity sector of an island pair is inconsequential to the qubit, and consider more general error sources that are not described by the processes discussed in the main text.
We define the three Pauli operators in the space of even and odd parity states $\{\ket{e},\ket{o}\}$ of a topological superconducting island
\begin{equation}
\tau_x = \begin{pmatrix}
0 & 1 \\
1 & 0
\end{pmatrix},\ \tau_y = \begin{pmatrix}
0 & -i \\
i & 0
\end{pmatrix},\ \tau_z = \begin{pmatrix}
1 & 0 \\
0 & -1
\end{pmatrix}\ .
\end{equation}
The four Majorana operators $\gamma_1,\dots, \gamma_4$ of an island pair can be represented in terms of these Pauli operators as
\begin{equation}\begin{split}
	\gamma_1 &= \tau_x \otimes \mathbbm{1} \, , \, \, \, \gamma_2 = -\tau_y \otimes \mathbbm{1} \, ,\\
	\gamma_3 &= \tau_z \otimes \tau_x \, , \, \gamma_4 = -\tau_z \otimes \tau_y \, ,
\label{eqn:majoranapauli}
\end{split}\end{equation}
upon choosing a specific order of modes and by invoking the Jordan-Wigner transformation.
These operators are Hermitian $\gamma_i = \gamma_i^\dagger$ and fulfill the anti-commutation relations $\{\gamma_i,\gamma_j\} = 2\delta_{i,j}$. Our two qubit encodings are
\begin{equation}
	\ket{0_e} = \ket{e} \otimes \ket{e} \, , \, \ket{1_e} = \ket{o} \otimes \ket{o}
\label{eqn:enc1}
\end{equation}
in the even parity sector and
\begin{equation}
	\ket{0_o} = \ket{e} \otimes \ket{o} \, , \, \ket{1_o} = \ket{o} \otimes \ket{e}
\label{eqn:enc2}
\end{equation}
in the odd parity sector. Therefore, in both parity sectors, the logical qubit operators are $\sigma_z = i\gamma_1 \gamma_2$ and $\sigma_x = i \gamma_2 \gamma_3$. Consider a quasiparticle poisoning event described by the application of $\gamma_1$. The operator $\gamma_1$ maps $\ket{0_e} \leftrightarrow \ket{1_o}$ and $\ket{1_e} \leftrightarrow \ket{0_o}$, i.e., it switches the parity sector and applies a logical $\sigma_x$ operation. Similarly, Eq.\ \eqref{eqn:majoranapauli} implies that $\gamma_2$ applies a logical $\sigma_y$ operation and $\gamma_3$ a logical $\sigma_z$ operation. The operator $\gamma_4$ only switches the parity sector without changing the logical information. This is not surprising, since it is the only Majorana operator that is not part of either $\sigma_z$ or $\sigma_x$. What is more, invoking the fermion-parity superselection rule, it is clear that the 
specific order of modes used in this argument is not relevant, i.e., that the specific Jordan-Wigner string plays no role.

\begin{figure*}
\centering
\def\svgwidth{\linewidth}
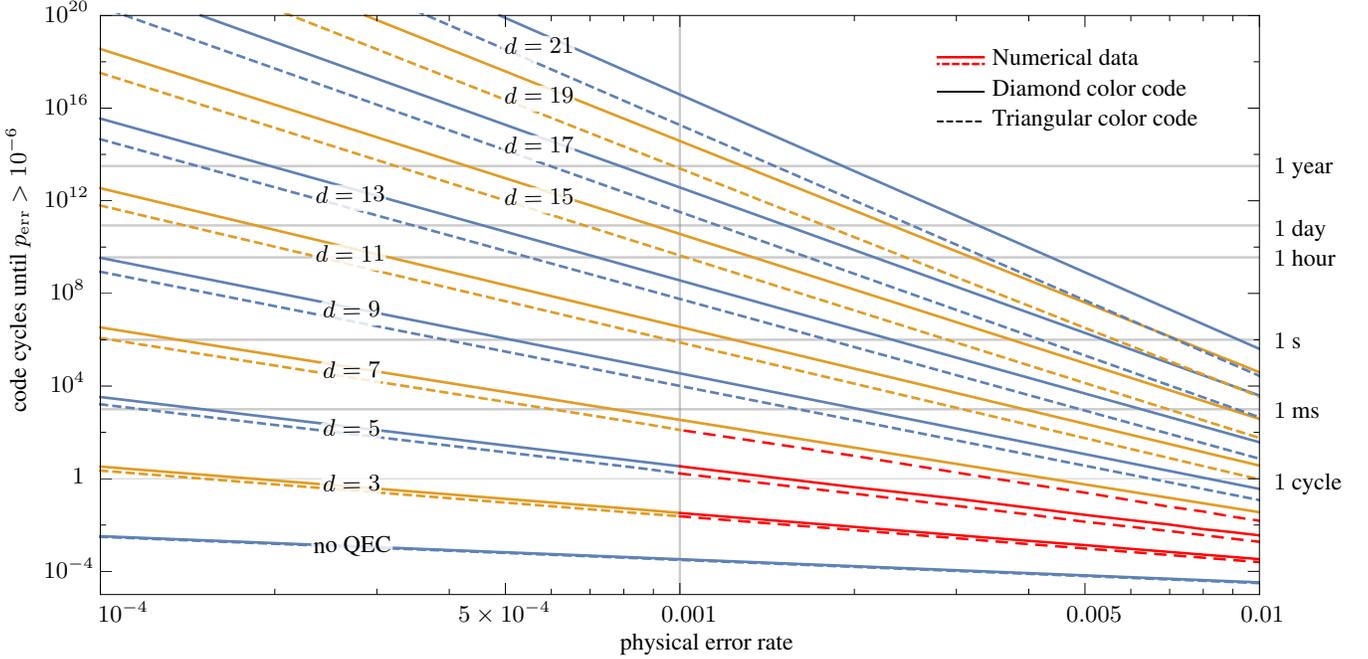
\caption{Survival time of triangular (dashed) and diamond (solid) color code qubits until the logical error probability 
$p_{\rm err}$ reaches $10^{-6}$ for increasing code distances obtained from extrapolation of the numerical data (red). The labels on the right-hand side of the time axis show the survival time for a code cycle duration of $1 \mu s$.}
\label{fig:extrapolation}
\end{figure*}

To further demonstrate that, in general, the information about the parity sector is irrelevant for quantum computation, we write the state of an island pair as a product of the \textit{qubit state} and the \textit{parity state}. An island pair is a four-level system with the four basis states in Eqs.\ \eqref{eqn:enc1} and \eqref{eqn:enc2}. Instead of describing these states in terms of the fermion parities of the first and second island, we can transform the basis to a product of an eigenstate $\{\ket{0},\ket{1}\}$ of the qubit operator $\sigma_z = i\gamma_1\gamma_2$ (qubit state) and an eigenstate $\{\ket{p_e},\ket{p_o}\}$ of the total parity operator $p=-\gamma_1\gamma_2\gamma_3\gamma_4$ (parity state). In this basis, the four states are
\begin{equation}\begin{split}
	\ket{0_e} = \ket{0} \otimes \ket{p_e} \, &, \, \ket{1_e} = \ket{1} \otimes \ket{p_e} \, , \\
	\ket{0_o} = \ket{0} \otimes \ket{p_o} \, &, \, \ket{1_o} = \ket{1} \otimes \ket{p_o} \, .
\label{eqn:enc3}
\end{split}\end{equation}
Incidentally, the transformation matrix which maps from \eqref{eqn:enc1} and \eqref{eqn:enc2} to \eqref{eqn:enc3} is a CNOT. Braiding operations and measurements of the qubit only affect the qubit state but not the parity state, since they are comprised of operators that are products of $\gamma_1\gamma_2$ or $\gamma_2\gamma_3$, and therefore commute with the parity operator $p$. After this mapping, the poisoning processes that we considered previously (i.e., the application of a Majorana operator) can be written as a product of operations on the qubit state and on the parity state
\begin{equation}\begin{split}
	\gamma_1 = \widetilde{\sigma}_x \otimes \widetilde{\tau}_x\, &, \, \gamma_2 = -\widetilde{\sigma}_y \otimes \widetilde{\tau}_x\, , \\ \gamma_3 = \widetilde{\sigma}_z \otimes \widetilde{\tau}_x\, &, \, \gamma_4 = -\mathbbm{1} \otimes \widetilde{\tau}_y\, ,
\end{split}\end{equation}
where $\widetilde{\sigma}_i$ and $\widetilde{\tau}_i$ are Pauli operators acting on the qubit and parity spaces, respectively.
Therefore, these operations do not entangle the qubit with the parity degree of freedom. Furthermore, any Jordan-Wigner string $\tau_z \otimes \tau_z$ associated with the operators in Eq.\ \eqref{eqn:majoranapauli} is mapped onto $\mathbbm{1} \otimes \widetilde{\tau}_z$, which acts trivially on the qubit state. However, a general operation can in principle generate such an entangled state. Thus, the most general description of the entire state of the system is a sum over all $2^n$ possible $2n$-island parity sectors
\begin{equation}
	\ket{\psi} = \sum\limits_{\mathrm{parity~sectors}~p} \ket{\psi_p} \otimes \ket{p}\, ,
\end{equation}
where $\ket{p}$ contains all fermion parities of the $n$ island pairs, and $\ket{\psi_p}$ is an $n$-qubit state. But in our encoding, $\ket{p}$ carries no information relevant to quantum computation. Tracing out the parity state leaves the qubit state in a statistical ensemble. Therefore, the parity degree of freedom acts like an environment, to which error sources can couple. Moreover, in the absence of logical errors, different qubit states yield different syndromes after stabilizer readout. Thus, measuring the syndrome breaks the entanglement between the qubit state and parity state.

Error sources that entangle the qubit with the parity state are not described by products of Majorana operators, but by sums of products. Such errors are in principle allowed and lead to a non-unitary evolution of the qubit state. These errors are still correctable, but are not necessarily described by the error model of random Pauli errors. One example for such an effectively non-unitary process is swapping the parities of two islands that belong to two different qubits, as this entangles the qubit and parity degrees of freedom.

\section{Monte Carlo simulation of the diamond color code}
\label{app:numerics}

In order to study the performance gain of low-distance diamond color codes over triangular color codes, and estimate the logical error rate for higher-distance codes, we sample the logical error rate in a Monte Carlo simulation. The physical error rate is the probability for at least one error event in a code cycle
\begin{equation}
	p_{\rm phys} = 1- \lim\limits_{N \rightarrow \infty} \left( 1- \frac{1}{N} \frac{\tau_c}{\tau_p} \right)^N = 1- e^{-\tau_c/\tau_p} \, ,
\end{equation}
where $\tau_c$ is the duration of a code cycles and $\tau_p$ is the characteristic time-scale on which bit flips and phase flips occur. A physical bit flip or phase flip only occurs at the end of a code cycle if the bit is flipped an odd number of times within a cycle. The probability of a physical bit flip or phase flip can be calculated from the probability of an odd number of successes in $n$ discrete trials with success probability $p$, which is
\begin{equation}
	p_{\rm odd} = \frac{1-(1-2p)^n}{2} \, .
\end{equation}
Thus, the physical bit flip (and phase flip) probability is
\begin{equation}\begin{split}
	p_{\rm flip} &= \lim\limits_{N \rightarrow \infty} \frac{1}{2} \left( 1- \left(1-\frac{2}{N} \frac{\tau_c}{\tau_p} \right)^N \right) \\
	&= \frac{1}{2}\left(1-e^{-2\tau_c/\tau_p}\right) = p_{\rm phys} - \frac{1}{2} p_{\rm phys}^2 \, .
\end{split}\end{equation}
We define the logical error rate $p_{\rm log}$ as the probability for a logical bit flip (or phase flip). Without quantum error correction, $p_{\rm log} \neq p_{\rm phys}$ since the absence of a logical error requires the absence of both $\sigma_x$ and $\sigma_z$ errors. Thus, the physical qubit needs to pass two trials and the logical error rate is
\begin{equation}
	p_{\rm log} = 1- \left(1- p_{\rm flip} \right)^2 \, .
\end{equation}
To calculate $p_{\rm log}$ with error correction, we sample through error configurations with a bit flip probability $p_{\rm flip}$ on each physical qubit, attempt to correct the error using a decoder, and count the number of failure events. The logical error rate is
\begin{equation}
	p_{\rm log} = 1 - \left( 1 - \frac{{\rm fails}}{{\rm trials}} \right)^2 \, .
\end{equation}
Our decoder is a pregenerated look-up table which, given an error syndrome, returns the most likely corresponding error configuration. 
Since this is not efficient for higher-distance codes, we only simulate the triangular color codes with distances $d=3$, $d=5$, and $d=7$, and diamond color codes with distances $d=3$ and $d=5$. On a log-log-plot, the logical error rate is linear for low physical error rates, see Fig.~\ref{fig:numerics}. The slopes and offsets of these linear functions both grow approximately linearly with increasing code distance, allowing for a rough estimate of the low-error behavior of higher-distance codes through extrapolation.

The survival time of a logical qubit until the probability of a logical error $p_{\rm err}$ reaches a target accuracy $p_{\rm target}$ is 
\begin{equation}
	\tau_{\rm survival} = \frac{\ln (1-p_{\rm target})}{\ln(1-p_{\rm log})} \, ,
\end{equation}
where $\tau_{\rm survival}$ is the survival time as a number of code cycles. In Fig.~\ref{fig:extrapolation}, we plot the survival time for $p_{\rm target} = 10^{-6}$ for triangular and diamond color codes obtained from numerical data and extrapolation thereof. The extrapolation indicates that for $p_{\rm phys} = 10^{-3}$ and a code cycle duration $\tau_c = 1 \mu s$, the survival time of a $d=19$ diamond color code is of the order of several years.

\section{4.8.8 vs 6.6.6 color codes}
\label{app:488comp}

Quantum error-correcting codes are usually classified using their code distances and error thresholds. For triangular and diamond color codes, we have seen that two 6.6.6 color codes with equal code distances can exhibit different logical error rates. Here, we show that neither the code distance nor the error threshold are predictive figures of merit for the logical error rate, which determines the performance of a code.

In this work, we considered color codes that are defined on lattices with 6.6.6-tiling. A different tiling that allows for color codes is the 4.8.8-tiling, with two types of eight-qubit stabilizers and one type of four-qubit stabilizers. These 4.8.8 codes are considered to be more efficient, since triangular 4.8.8 codes require fewer physical qubits per logical qubit compared to triangular 6.6.6 code with the same code distance.

\begin{figure}[t]
\centering
\def\svgwidth{\linewidth}
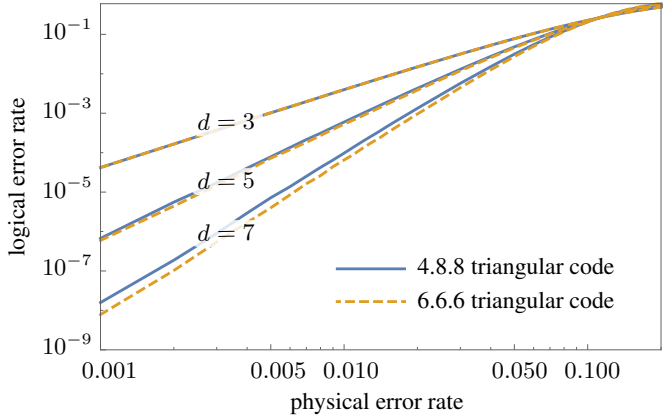
\caption{Logical error rate of the first three 4.8.8 and 6.6.6 triangular color codes. Even though the codes have the same code distance and error threshold, 4.8.8 color codes have a higher logical error rate, but require fewer physical qubits per logical qubit.}
\label{fig:488comp}
\end{figure}
Figure \ref{fig:488comp} shows the logical error rate of 4.8.8 and 6.6.6 codes obtained from the previously described Monte Carlo simulation. It shows that the lower physical overhead of 4.8.8 codes comes at the price of a higher logical error rate. Therefore, for a target logical error rate, it is difficult to estimate which code one should use to minimize the physical overhead. To our knowledge, even though code distances and error thresholds of codes are well-studied, logical error rates have so far attracted less attention, and require further research.

\bibliographystyle{apsrev4-1mod}
\bibliography{biblio}

\end{document}